\documentclass[a4paper,twocolumn,11pt,accepted=2025-07-01]{quantumarticle}
\pdfoutput=1
\usepackage{adjustbox}

\usepackage[symbol]{footmisc}
\usepackage{amsmath,amsthm,amssymb}
\usepackage{graphicx} 
\usepackage{mathtools}
\usepackage{epstopdf}
\usepackage{tabularx}
\usepackage{tabularray}
\usepackage[normalem]{ulem}
\usepackage[T1]{fontenc}

\usepackage{float}
\makeatletter
\let\newfloat\newfloat@ltx
\makeatother

\usepackage{algorithm}
\usepackage{algpseudocode}
\usepackage{braket}
\usepackage{tikz}
\usetikzlibrary{quantikz2}

\DeclareGraphicsExtensions{.png,.pdf}
\usepackage{tabularx}

\usepackage{xcolor}
\usepackage{hyperref}
\usepackage{graphicx}
\usepackage{dcolumn}
\usepackage{cleveref}
\usepackage{dutchcal}
\usepackage{nag}
\usepackage{graphicx}
\usepackage{amsthm}
\usepackage{tabulary}
\usepackage{mathtools}
\usepackage{xcolor}
\usepackage{braket}
\usepackage{datetime}
\usepackage{stackrel}
\usepackage{stmaryrd}
\usepackage{dsfont}
\usepackage{units}
\usepackage{float}
\usepackage{relsize}
\usepackage{bbm}
\usepackage{tikz}

\newcommand{\negspace}{\!}

\newcommand{\rsub}[2]{{#1} \negspace {\protect\vphantom{#1}}_{#2}}

\newcommand{\ketsub}[2]{\rsub {\ket{#1}} {#2}}

\begin{document}

\title{End-to-end switchless architecture for fault-tolerant photonic quantum computing}

\author{Paul Renault}
\thanks{These authors contributed equally to this work}
\author{Patrick Yard}
\thanks{These authors contributed equally to this work}
\author{Raphael C.~Pooser}
\thanks{These authors contributed equally to this work}
\author{Miller Eaton}
\author{Hussain Asim Zaidi}
\email{hussain@qc82.tech}

\address{QC82 Inc., 7757 Baltimore Ave, College Park, MD 20740}

\begin{abstract}
Photonics presents one of the most promising approaches to large-scale quantum computation with millions of qubits and billions of gates, owing to the potential for room-temperature operation, high clock speeds, miniaturization of photonic circuits, and repeatable fabrication processes in commercial photonic foundries. 
We present an end-to-end architecture for fault-tolerant continuous-variable (CV) quantum computation using only passive on-chip components that can produce photonic qubits above the fault tolerance threshold with probabilities above 90\%, and encodes logical qubits using physical qubits sampled from a distribution around the fault tolerance threshold. 
By requiring only low photon number resolution, the architecture enables the use of high-bandwidth photodetectors in CV quantum computing.
Simulations of our qubit generation and logical encoding processes show a Gaussian cluster squeezing threshold of 12~dB to 13~dB. 
Additionally, we present a novel magic state generation protocol which requires only 13~dB of cluster squeezing to produce magic states with an order of magnitude higher probability than existing approaches, opening up the path to universal fault-tolerant quantum computation at less than 13~dB of cluster squeezing.

\end{abstract}


\maketitle

\section{Introduction}\label{sec:Intro}
Quantum computation promises the ability to exponentially speed up certain problems intractable to classical computers including simulations of large quantum systems, an observation first formalized by Richard Feynman~\cite{feynmanSimulatingPhysicsComputers1982}. Noisy intermediate scale quantum computers~\cite{Preskill2018} with $\approx 100$ physical qubits and $\approx 100$ to $1000$ quantum gates have been demonstrated on a variety of physical platforms~\cite{debnathDemonstrationSmallProgrammable2016a,kandalaHardwareefficientVariationalQuantum2017,henrietQuantumComputingNeutral2020,omalleyScalableQuantumSimulation2016}, but it is widely expected that the computers must scale to millions of physical qubits and billions of operations to be useful for commercially valuable quantum simulations~\cite{kim2022fault,gidney2021factor,litinski2023compute}.
Matter based approaches face hurdles to get to such a computational volume, owing to the hardware requirements of low temperatures, slow gate rates, or lossy conversions between matter and photons~\cite{meesalaQuantumEntanglementOptical2024,kandalaHardwareefficientVariationalQuantum2017,omalleyScalableQuantumSimulation2016,henrietQuantumComputingNeutral2020,saffmanQuantumComputingAtomic2016}. 
Photonics provides a promising path towards large-scale quantum computing with scalable generation and manipulation of qubits at room temperature, high-speed measurements, miniaturization using photonic integrated circuits, and repeatable low-loss mass manufacturing at commercial foundries. 

Photonic quantum information can be encoded into either discrete quantum degrees of freedom of individual single photons, known as the discrete variable (DV) approach \cite{Kok2005,Bartolucci2021}, or in continuous-valued orthogonal quadratures of the electric field, known as the continuous variable (CV) approach~\cite{Lloyd1999,Fabre2020}.
CV quantum photonics brings an advantage over DV approaches, namely the deterministic generation of entanglement over large photonic quantum states~\cite{Weedbrook2012,Menicucci2014ft}. Millions of entangled optical modes have been experimentally generated in CV systems~\cite{Asavanant2019}. 
This makes CV photonics ideal for measurement-based QC (MBQC) implementations~\cite{Menicucci2006}, in which quantum computation is performed via single-qubit measurements on a large entangled quantum state. In photonics, single modes or collections of modes in the entangled state act as the analogues of qubits in matter-based approaches.

The key concept to enable quantum computation with large-scale entangled states, or in general with a large number of qubits, is that of fault tolerance. \emph{Fault-tolerant quantum computing} (FTQC) is an approach to QC in which a computational (logical) qubit is encoded in many physical qubits on which quantum error correction (QEC) protocols are employed to correct for errors such as decoherence and photon loss~\cite{Gottesman2001,caiBosonicQuantumError2021,xuAutonomousQuantumError2023};
If the probability of physical errors is below a certain value determined by the QEC protocol, called the \emph{fault tolerance threshold}, the probability of logical error can be exponentially suppressed. There exist no direct fault-tolerant protocols for quadrature encoded continuous-variable quantum computing (CVQC)~\cite{ohligerLimitationsQuantumComputing2010a}. An alternative approach is to discretise the CV space into qubits~\cite{ohligerLimitationsQuantumComputing2010a, Menicucci2014ft}.

The most promising qubits for CV FTQC are the \emph{Gottesman-Kitaev-Preskill} (GKP) qubit~\cite{Gottesman2001} and the \emph{magic states}~\cite{Bravyi2005}. Native logical gates are implemented with Gaussian operations on GKP qubits, while universal quantum computation is possible with the ability to create magic states. 
In fact, using only CV Gaussian resources combined with GKP states, one has all the necessary ingredients to achieve universal CVQC~\cite{Baragiola2019}. 
Therefore, much effort in CVQC has gone into devising scalable methods of producing high quality GKP qubits (that is, with high effective squeezing; see Sec.~\ref{sec:CVtools}), but producing these states with high success rates in photonics has remained a challenge (see Sec.~\ref{sec:GKP}). 
After obtaining high quality GKP and magic qubits, CV architectures achieve fault tolerance by concatenating a second qubit-level error correcting code with the GKP states as the physical qubits. To implement these codes with optical fields, photonic switches were proposed either to increase the low success probabilities of GKP state generation~\cite{Bourassa2021blueprintscalable, tzitrin2021fault} or for quantum error correction~\cite{larsen2021fault} (see Sec.~\ref{thresh_sim}). Great progress has been made in lowering the noise and losses in photonic switches, but the loss levels are two orders-of-magnitude higher than passive photonics~\cite{psiquantum2025manufacturable}, making removal of switches from the architecture a desirable goal.

With this context, we present four contributions in synthesizing an end-to-end CV FTQC architecture. First, we provide a protocol that produces fault-tolerant CV photonic qubits --GKP qubits-- at state-of-the-art probabilities of more than $90\%$. 
Notably, the architecture makes use of all physical qubits, both above and below the FTQC threshold, to construct logical qubits. 
Second, we provide methods to generate GKP qubits on an integrated switchless photonic chip, opening up the possibility of low-loss manufacturability of the photonic chips with passive components. 
Our methods for state generation require only low photon number resolution (PNR) for which either superconducting microstrip single photon detectors (SMSPDs)~\cite{kong2024} or room temperature photon number resolving detectors~\cite{Nehra2020b} would suffice. 
Third, we present protocols for magic state generation with state-of-the-art success probability, an order of magnitude better than existing methods, for universal FTQC with feasible resource requirements. 
Fourth, we present the results of our quantum error correction simulations including GKP resource generation costs. We arrive at the \emph{gaussian squeezing} threshold, the quantum noise metric in CVQC~\cite{Menicucci2014ft, schnabel20171}, of $12$ dB to $13$ dB for fault tolerance, and distillable magic state success probability of $4.8\%$ also at $13$ dB. 
To our knowledge, the presented simulations are the first end-to-end CVQC simulations incorporating a realistic process producing GKP states with a range of effective squeezing values for fault-tolerant operation. 

In the following sections, we start by providing an overview of photonic CVQC. We then describe our architecture before proceeding to discussing the details and physics of each architectural component, and conclude by providing the results of the fault tolerance simulations. 

\section{CVQC Toolkit and Non-Gaussian States}\label{sec:CVtools}
In photonic CV MBQC, the set of entangling operations, non-destructive gates, and destructive measurements come from the CV photonics toolkit~\cite{Pfister2019}, the most prominent components of which are the squeezing operations, linear optics, quadrature displacements, homodyne measurements, and photon number resolving measurements. 
In the following, we define the key terms, including entangled states, measurement operations, and non-Gaussian modes, which form the basic building blocks of our architecture.

The entangled state often used in photonic CV MBQC is the \emph{cluster state}, canonically defined as Gaussian-squeezed modes entangled using controlled-phase gates ($C_Z$)~\cite{nielsen2006cluster}. While CVQC was initially investigated with Gaussian cluster states requiring inline squeezing due to the use of controlled-phase gates \cite{Raussendorf2001,Flammia2009,Menicucci2006}, an alternative formalism exists where inline squeezing is replaced by offline squeezing and passive interferometers \cite{Menicucci2011a,van2007building,walshe2021streamlined}, resulting in a \emph{macronode cluster state}, where collections of physical optical modes can be treated as a single mode on an underlying lattice~\cite{Alexander2016a}. An example of macronode cluster is the multimode state called a dual rail quantum wire,  generated using the two-mode-squeezing operation implemented in a non-linear crystal followed by passive beamsplitters~\cite{Pysher2009}. 

The name squeezing stems from the shape of the Wigner quasiprobability distribution of the conjugate photonic quadratures (denoted $p$ and $q$), in which the distribution for one quadrature has a narrower width compared to a coherent state, while the other quadrature has a broader distribution. 
The term also refers to the quantum noise on a measurement along one or a combination of the quadrature axes in phase space. In CV MBQC, because all gates are carried out via measurement operations, the noise on these gates is parametrized by the squeezing observed in each measurement. Thus, physical gate noise is directly tied to the squeezing level associated with each mode being measured.
Ideal Gaussian squeezed states (infinitely squeezed) are denoted by $\ket{m}_p$, where $m$ is the eigenvalue of the quadrature $p$. For the results presented in this paper, all simulations are run with finitely squeezed states (i.e., resulting from applying the finite squeezing (single-mode or two-mode) operator to vacuum). 
FT thresholds in terms of squeezing are important for CV FTQC as single and two-mode squeezing forms the fundamental resource for CVQC~\cite{guQuantumComputingContinuousvariable2009,Menicucci2007}.

The operations used to project a quantum state onto quadratures eigenstates are called \emph{homodyne measurements}~\cite{lvovsky2009}: a homodyne measurement with outcome $m$ on quadrature $p$ projects the measured mode onto the state $\ket{m}_p$. An important property of homodyne measurements chosen with appropriate bases is that they can be used to either delete a node from a cluster state or teleport the quantum information from one node to another~\cite{guQuantumComputingContinuousvariable2009}, allowing one to shape cluster states during quantum computation. 

While deterministic and scalable Gaussian entanglement generation gives us a substrate quantum state to use in CVQC, and homodyne measurements can be used to shape the cluster state and implement quantum gates, universal CVQC requires additional resources to enable quantum computations that are not efficiently classically simulable~\cite{Bartlett2002}. One prominent operation to enable universal CVQC is photon subtraction~\cite{ourjoumtsev2006generating}, which can be used to implement gates consisting of arbitrary powers of the quadrature operators on squeezed states. Experimentally, in quantum optics, the most common method for performing photon subtraction involves using a highly reflective beamsplitter, with a photon detector—capable or not of resolving the number of detected photons— placed in the transmission arm \cite{ourjoumtsev2006generating,melalkia2022plug}. Throughout this article, whenever the photon subtraction process is mentioned, we refer to this experimental setup.
One of the key applications of these higher-order gates is the production of \emph{non-Gaussian states} in an entangled cluster, which along with homodyne detections and displacements, give us the ability to apply quantum error correction and operate the universal quantum computer above the FT threshold~\cite{Menicucci2014ft}. 

\begin{figure*}
    \centering
    \includegraphics[width=0.25\textwidth]{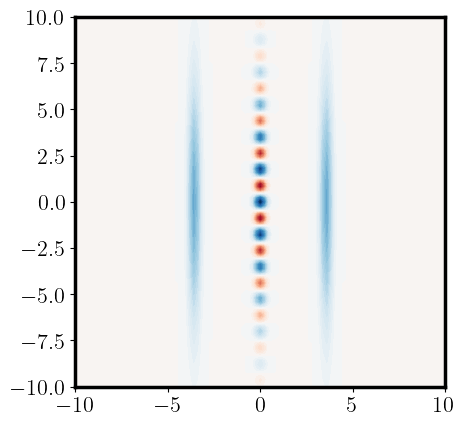} 
    \includegraphics[width=0.25\textwidth]{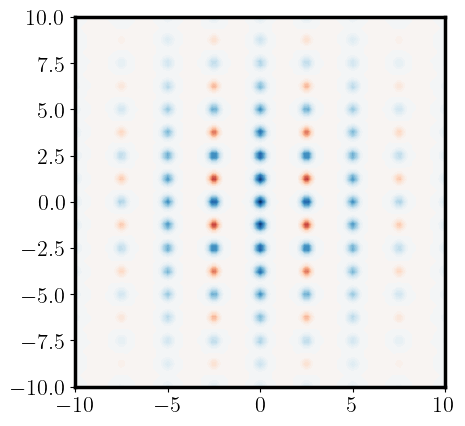} \includegraphics[width=0.25\textwidth]{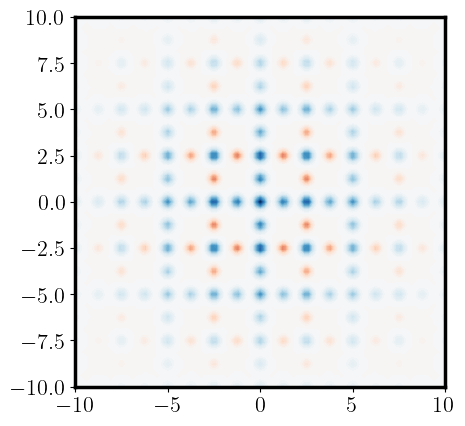}
    \caption{Wigner functions of squeezed cat state, GKP zero qubit $\ket{0}_G$, and magic H state $\ket{H}$. The horizontal and vertical axes are the $q$ and $p$ quadrature axes, respectively. Each of the states has squeezing of approximately 13dB in the $q$ direction (cat state) or both the $q$ and $p$ directions (GKP and magic states). Red peaks correspond to negative values of the Wigner distribution, while blue peaks are positive. The color scale is otherwise arbitrary and is meant only to notionally represent the location of probability density.}
    \label{fig:gridWigs}
\end{figure*}

The fundamental non-Gaussian resource states in our architecture are the squeezed cat states $\ket{\pm \alpha, r}$, which are superpositions of displaced squeezed states, i.e., $\ket{\pm \alpha, r}\sim (D(\alpha) + D(-\alpha))\mathcal{S}(r)\ket{0}$, where $D$ is the displacement operator in the phase space, $D(\alpha)=\exp{(\alpha a^\dagger -\alpha^\star a)}$, $\mathcal{S}$ is the squeezing operator, $\mathcal{S}(r)=\exp(\frac{1}{2}(r^*a^2 - ra^{\dagger^2}))$, and $a^\dagger$ and $a$ are the creation and annihilation operators, respectively. Real $\alpha$ results in $q$ quadrature displacements ($D_q$), while imaginary values result in $p$ displacements ($D_p$). 
Cat states are parametrized by squeezing ($r$) and displacement ($\alpha$), which we take to be real-valued in the following results. 
The Wigner function for a squeezed cat state is shown in Fig.~\ref{fig:gridWigs}. 
While FTQC codes exist with cat states as the logical qubits, these require the experimentally-challenging parity check operator~\cite{Schlegelcatqec2022}, limiting the role of cat states in photonic FTQC to that of resource states which we consume to produce qubits.

GKP states, and more generally grid states, were first proposed in [Ref.~\cite{Gottesman2001}], with a considerable amount of effort spent studying their use in fault-tolerant architectures in recent years~\cite{tzitrin2020progress, mensen2021phase, grimsmo2021quantum}; a typical Wigner function is shown in~\ref{fig:gridWigs}. 
The sensor state, denoted $\ket{S}$, is a symmetric grid state defined as the $+1$ eigenstate of the displacements $D_q(\sqrt{2\pi})$ and $D_p(\sqrt{2\pi})$, while a GKP qubit is the $+1$ eigenstate of the displacements $D_q(2\sqrt{\pi})$ and $D_p(\sqrt{\pi})$, where we have set $[a,a^\dagger]=1$
~\cite{Gottesman2001, grimsmo2021quantum} (see Appendix Sec.~\ref{appterm} for more details).
Interference between sensor states using passive linear optics can be used to produce entangled GKP Bell pairs ~\cite{walshe2020continuous}.

Ideal GKP states have wavefunctions that are a series of delta functions in the $q$-basis (or $p$-basis) of the continuous-variable $(q,p)$ phase space of a bosonic mode, and are therefore unphysical (see Appendix \ref{appterm} for ideal GKP wavefunctions). 
Physically-relevant approximations to the ideal GKP states can be obtained by applying the damping operator to the ideal state~\cite{Menicucci2014ft,matsuura2020equivalence}. 
The width of the peaks provides an \textit{effective squeezing} parameter, which describes the closeness of a given GKP state to the ideal state. 
The probability of physical gate errors in CVQC depends on the effective squeezing of the underlying GKP states, which denotes the noise level of measurements associated with the qubit compared to the standard noise of a coherent state~\cite{Weigand2018}; minimizing this noise corresponds to reduced physical error rates.
For any given density matrix $\rho$ and the desired displacement operator with displacement $u$, we estimate the effective squeezing of the output state as $\Delta_D = \frac{1}{u}\sqrt{\ln{|Tr D(u)\rho|}}$ where $Tr$ is the trace operation~\cite{Weigand2018}. 
The quality of a realistic GKP state for FTQC is quantified most readily by the effective squeezing parameter since the probability of error in the logical X or Z operations is bounded by $P < \frac{2\Delta}{\pi}\exp{-\pi/(4\Delta^2)}$, where $\Delta$ is the effective squeezing in the respective quadrature~\cite{Weigand2018}. 
Note that the GKP state can have different effective squeezing in the two quadratures owing to different parameters at play in the generation protocols. 
A final important non-Gaussian state in our architecture is the magic state. These are essential for universal quantum computation by providing an implementation of a complete gate set.
In CVQC, magic states are also grid states; their Wigner functions contain periodic structure (see Fig.~\ref{fig:gridWigs}; also Appendix Sec.~\ref{appterm}), and they can be derived from CV GKP states~\cite{Baragiola2019}.
An important metric is the distillation threshold of magic states, which quantifies the fidelity above which many copies of noisy magic states can be used to distill a high fidelity magic state useful for applying fault-tolerant non-Clifford gates, e.g., the $T$ gate~\cite{Bravyi2005}. 

\section {Architecture and Design Overview}
\begin{figure*}
    \centering
    \includegraphics[width = 0.9\textwidth]{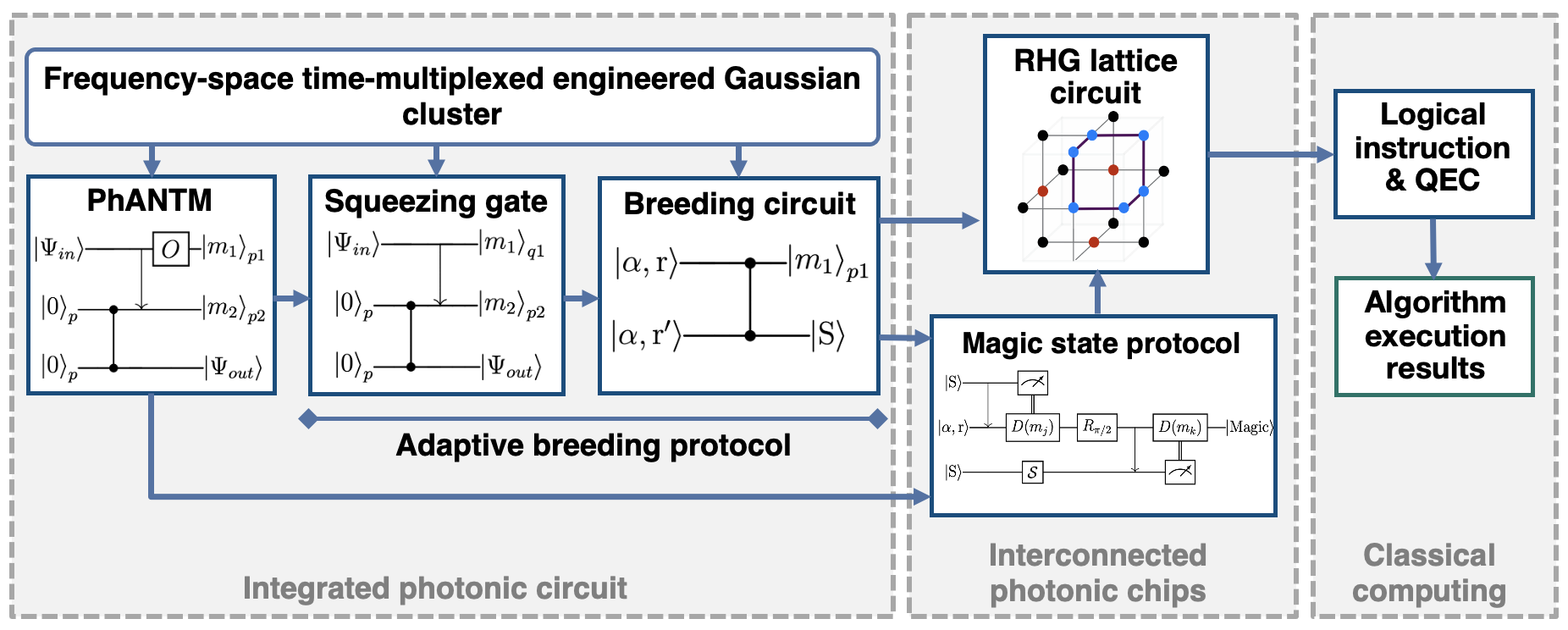}
\caption{Conceptual overview of the creation of GKP states, magic states, MBQC-based lattices, and QEC to enable FTQC. Repeating photon subtraction and teleportation along a cluster state generates cat states with a high success probability and low photon number resolution. The cats are then squeezed to produce cats with deterministic amplitudes which are bred together to create GKP states without the need for post-selected measurements. All steps happen on the cluster state. The GKP states and cat states are used together to create magic states. The architecture produces $H$-type magic states in this configuration. 
Notations used in circuits are defined in main text. Additionally, $\mathcal{S}$ and $R$ stand respectively for squeezing and rotation operators. Note that for brevity we use the notation for infinitely squeezed Gaussian states in the circuits, but in practice they are replaced by finitely squeezed states in the simulations.  Downward arrows in circuits are beamsplitters while vertical lines with circular caps are $C_Z$ gates.
}
    \label{fig:pipeline}
\end{figure*}

Our photonic QC architecture contains three key components for physical qubit generation (see Fig~\ref{fig:pipeline}): cat state generation via the Photon-counting-Assisted Node-Teleportation Method (PhANTM)~\cite{Eaton2022Phantm} (see also section~\ref{sec:phantm}), an adaptive breeding protocol (detailed in section~\ref{sec:GKP}), and a magic state generation protocol (described in section~\ref{sec:magic}). 
The logical layer consists of two major components: logical qubit generation, which consists of distributing physical qubits as the nodes within the Raussendorf-Harrington-Goyal (RHG) lattice~\cite{Raussendorf2006}; and the quantum error correction (QEC) encoding/decoding system, which is also responsible for running logical instructions.
All of the physical qubit generation and RHG lattice generation components are possible on a set of integrated photonic chips without active photonic switches or variable beamsplitters on the quantum modes to minimize loss and noise. 
While fast electro-optic modulation is needed on local oscillators during homodyne measurements, the local oscillator power can be corrected to account for any losses. 

We use cluster engineering with time multiplexing and photon-number-resolving detection to prepare cat states, followed by adaptive breeding to create GKP states using homodyne measurements. 
Magic states are produced by applying CV QEC on cat states, using bred GKP states as resource states. 
We generate the RHG lattice using static linear optics and perform syndrome extraction and logical operations by single qubit homodyne measurements on logical qubits within the surface code. 
Classical compute resources are used to implement the user interface, code compilation, and transpilation to local oscillator phase controls, which implement all gate controls in the architecture. 
Homodyne detection measurement results are sent to the logical instruction and QEC system where syndrome decoding and error tracking are performed.
Any necessary modifications to the measurement results as a consequence of error correction are applied before using measurement outcomes to compute the algorithm results. 
When online changes to future measurements are required, the result of error detection is sent to the phase modulators which control the local oscillator phases that set the measurement bases.

The substrate to create GKP and magic states is a cluster state formed from two-mode squeezed light and passive beamsplitters, where each node is a photonic mode characterized by the quadratures $q$ and $p$, and features nearest-neighbor entanglement between the nodes. The time-multiplexed cluster state in our architecture is indexed by spatial and optical frequencies along with temporal modes (see Fig.~\ref{fig:cluster_ph} and later figures on cluster state engineering). To produce cat states our architecture uses PNR detection at specific optical frequencies and timesteps. Specifically, Schr\"{o}dinger cat states are produced using the Photon-counting-Assisted Node- Teleportation Method (PhANTM) \cite{Eaton2022Phantm}. PhANTM is well-suited to our time-frequency-space multiplexed architecture, as individual nodes of a cluster state can be made non-Gaussian with photon subtraction, and the non-Gaussianity can be teleported through a cluster state with homodyne measurements. This allows us to repeat the PhANTM process over multiple time steps to increase the amplitude of the generated cats and make them suitable for high-probability GKP generation, in contrast to single-shot GKP generation methods without time or frequency multiplexing~\cite{Bourassa2021blueprintscalable,Takase2024}. 

The cat states, which are embedded in the cluster state, are squeezed using teleportation-based squeezing gates~\cite{Alexander2014} and then ``bred'' into either sensor states for qubit production or GKP states for magic state production~\cite{Baragiola2019}. Switching between the two output states is achieved by changing the required amplitude of the input cat states.
Given that our protocols produce cat states with probabilistic squeezing and amplitude located at the nodes of a cluster state, we use an \emph{adaptive breeding} protocol to produce GKP qubits or sensor states. In our adaptive protocol, cat states of various multiples of the desired grid state spacing are used including a single mode squeezed vacuum that can be substituted for lower quality cat states when desired (see Algorithm ~\ref{GKPalg}).

\begin{figure*}
    \centering
    \includegraphics[width = 0.7\textwidth]{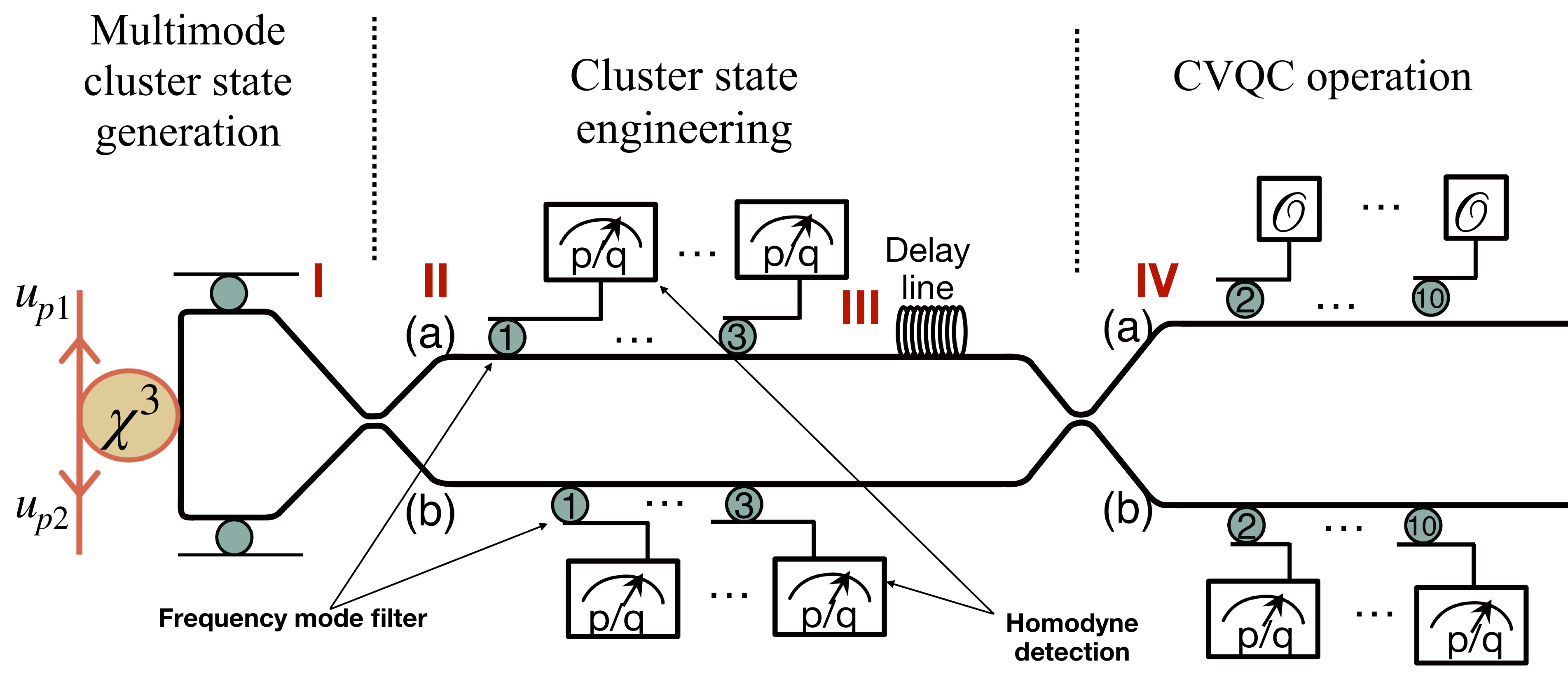}
\caption{Chip design of GKP generator to create multiple cat states in parallel from a time-frequency encoded Gaussian cluster state. The required squeezing is produced via four wave mixing in a ring resonator ($\chi^3$). Cat states are bred within the same chip to generate GKP state. The shape of the cluster state can be adapted as function of the operation applied in stage \textbf{IV}. $u_{p1}$ and $u_{p2}$ stand  for the optical pump beams at frequency respectively $\nu_{p1}$ and $\nu_{p1}$. Boxes with $p/q$ measurement indicate circuits where homodyne detection and corrective displacement are performed while boxes denoted $\mathcal{O}$ represent circuit where multiple photons subtraction are applied as well as displacement and homodyne detection. The green circles denote frequency filters that transmit given frequency modes. The labels within the circles correspond to the frequencies mode number consistent with those indicated in Fig. \ref{fig:cluster_ph}.}
    \label{fig:chip_v2}
\end{figure*}

These fundamental components can all be implemented in passive photonic integrated circuits, with squeezed light, sensor states, magic states, and the RHG lattice all produced on integrated chips.
The chips integrate linear optics components, such as beamsplitters and phase shifters, and nonlinear components such as $\chi^{(3)}$ media, as shown in Fig.~\ref{fig:chip_v2}. 
A squeezing resonator is bidirectionally pumped by pumps centered at $\nu_0$ with $\nu_{p_1}=\nu_0+\Delta \nu / 2$ and $\nu_{p_2}=\nu_0-\Delta \nu/2$, where $\Delta\nu$ is the free spectral range (FSR) of the resonator.
This creates two squeezed combs each centered about the respective pump. 
Interfering the two squeezed combs at a balanced beamsplitter results in the formation of a 1D dual-rail quantum wire \cite{Alexander2016b}. 
Subsequently, homodyne measurement are performed locally within this cluster to reshape it in preparation to apply PhANTM or adaptive breeding at different time steps. If the local oscillator phase is aligned with the $q$ quadrature then the node is removed whereas if the phase is that of the $p$ quadrature, the quantum information is teleported to a neighboring node, with an additional $\pi/2$ rotation applied \cite{Menicucci2006}. 
Each frequency mode needs to be separated from the optical frequency comb using frequency filters to then be detected. 
These filters are schematically shown as ring resonators, which are among the most compact and lowest-loss options for frequency separation~\cite{liuHighyieldWaferscaleFabrication2021} negligibly impacting the pass-through frequency modes, and demonstrating very high transmission efficiencies of the filtered mode.
The delay is used for time multiplexing before a final balanced beamsplitter is used to interfere frequency modes in (b) with the delayed modes in (a) to produce the final cluster state. With this resource state CVQC operations can be applied to generate cat states by performing PhANTM, squeezed cats using a squeezing gate, or GKP states by adaptive breeding. 

The architecture implements quantum error correction (QEC) by embedding the bred GKP sensor states from the photonic integrated circuit in Fig.~\ref{fig:chip_v2} into the RHG cluster state~\cite{Raussendorf2006}, the MBQC analog of the surface code. Large scale entanglement is generated using a macronode based approach~\cite{tzitrin2021fault} where each qubit of the RHG lattice is replaced by four optical modes. Entanglement links in the RHG lattice are replaced by GKP Bell pairs and the final cluster state is then generated by interfering all modes of the macronode with passive linear optics.

Logical gates can be implemented on the RHG cluster state using lattice surgery~\cite{herr2018lattice} and magic states. Magic states can be embedded in the macronode RHG lattice by replacing a GKP Bell pair with the result of interfering a magic state with squeezed vacuum~\cite{tzitrin2021fault}. More detail on the lattice construction is given in Sec.~\ref{logical}.

\section{Cat States on Dual-Rail Quantum Wire}\label{sec:phantm}
Cat states are a fundamental resource in our CVQC architecture. They are used as a resource state to generate both sensor states via breeding and magic states via implementing CV-QEC on input cat states.  Methods for creating cat states in CV photonics need at a minimum a combination of Gaussian and non-Gaussian resource states, or photon number detection along with Gaussian resource states. A method was proposed to create cat states by applying PNR on squeezed vacuum using machine learning to control some of the experimental parameters~\cite{Anteneh2024}. This method is efficient for state generation, requiring only a few iterations of photon detection, but requires high photon number resolution. Another proposal used high-photon-number Fock states (with $10$ or more photons) along with photon number resolution to create cat states deterministically~\cite{Winnel2024}. Creating high photon-number Fock states presents a high experimental challenge~\cite{uriaDeterministicGenerationLarge2020}. In PhANTM~\cite{Eaton2022Phantm}, the method used in the present architecture, applications of probabilistic photon subtraction and teleportation build up non-Gaussianity over time. This requires a Gaussian cluster as a resource to repeatedly apply the method. One of the key advantages of PhANTM lies in its ability to generate sufficiently large cat states that can subsequently be bred while distributing the total number of photons subtracted over multiple repetitions of the operation. 

\begin{figure*}
    \centering
    \includegraphics[width = 0.9\textwidth]{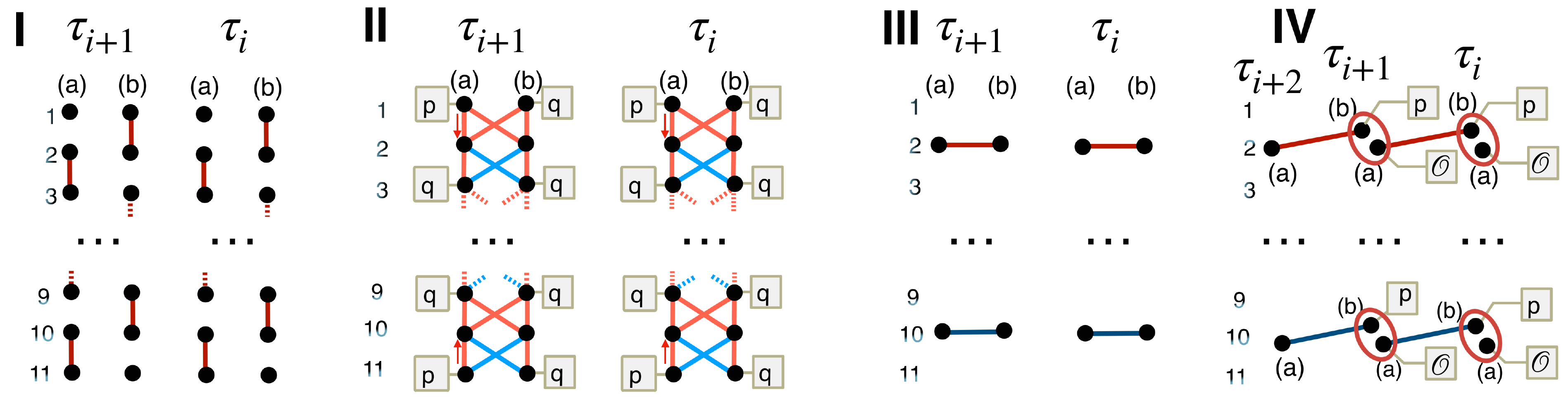}
    \caption{Cluster preparation for PhANTM application and parallel dual-rails. \textbf{I}, \textbf{II}, \textbf{III} and \textbf{IV} refer to locations on the chip in Fig. \ref{fig:chip_v2}. In stages \textbf{I} to \textbf{III}, the cluster state is generated and then reshaped using homodyne measurements to form dual rails independent of each other. In \textbf{IV}, PhANTM is applied as the circuit in Fig. \ref{fig:phantm_macronode} shows: at the time step $\tau_i$ modes $\nu_a$ and $\nu_{b}$ are respectively mode 2 and 1 in the circuit. Then the output mode in the dual rail is the mode $\nu_{b}$ of the $\tau_{i+1}$ time step. Dark blue and maroon segment represent effective $C_Z$ gates with weights of respectively $-1$ and $+1$, coming from applications of two-mode-squeezing and beamsplitter operations, while red circles represent the second beamsplitters in Fig.~\ref{fig:chip_v2}, where a phase can be applied to correct any rotation induced by cluster engineering. Orange and blue segments represent $C_Z$ gates with respective weights of $+1/2$ and $-1/2$. The red arrows in \textbf{II} show the direction of the teleportation.}
    \label{fig:cluster_ph}
\end{figure*} 

Fig.~\ref{fig:cluster_ph} shows the steps and cluster engineering using homodyne measurements, leading up to many copies of cats generated in parallel from a single frequency-space-time entangled cluster state. The frequencies are denoted by $\nu$ or arabic numerals, spatial rails are denoted by (a) and (b), and time steps are denoted by $\tau_i$. Starting with many pairs of two-mode entangled states in \textbf{I} interfered at the first beamsplitter of Fig.~\ref{fig:chip_v2}, we arrive at the entangled frequency chain also known as the dual-rail quantum wire shown in Fig.~\ref{fig:cluster_ph} \textbf{II}. We apply $p$ and $q$ homodyne measurements in region \textbf{II} of Fig.~\ref{fig:chip_v2} and Fig.~\ref{fig:cluster_ph} to convert a dual rail wire into multiple wires in frequency-space as shown in region \textbf{III} of Fig.~\ref{fig:cluster_ph}. Going through a time delay then entangles the two-mode wires in time to create many dual-rail wires in frequency-space-time as shown in Fig.~\ref{fig:cluster_ph}, on which PhANTM can be applied in parallel, e.g., on modes $2_{a,b}$ and $10_{a,b}$ in Fig. \ref{fig:cluster_ph} at time $\tau_{i}$. Since the cluster state modes that undergo homodyne detections or PhANTM are pre-determined, no active switching is required. 

\begin{figure}
    \centering
    \includegraphics[width = 0.3\textwidth]{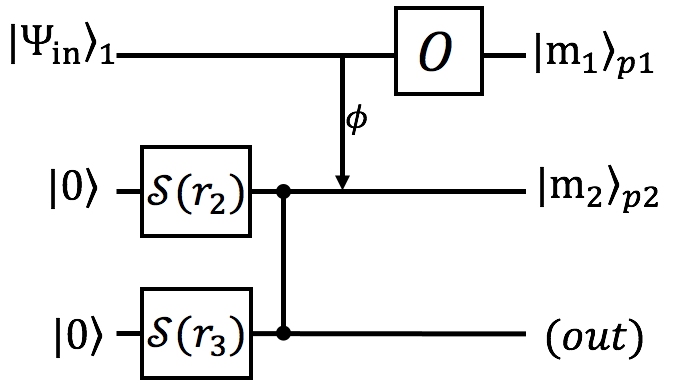}
    \caption{Application of PhANTM at stage \textbf{IV}. The top two modes are entangled through a beamsplitter (downward arrow) while the bottom two are entangled through a $C_Z$ gate (a line with two nodes at the ends) which is implemented by two-mode squeezing, beamsplitter, and homodyne measurements for cluster engineering. Conceptually, the $C_Z$ gate happens first, followed by the beamsplitter, and then the measurements from top to bottom. $O$ stands for multiple photon subtraction operations.}\label{fig:phantm_macronode}
\end{figure}

The application of PhANTM on the dual-rail wire in Fig.~\ref{fig:cluster_ph} \textbf{IV} corresponds to repeating the circuit in Fig.~\ref{fig:phantm_macronode} in time. 
The rotation on the output mode can be manipulated through the phase of the beamsplitter and homodyne detection basis. 
The beamsplitter is defined by two parameters, $\theta$ and $\phi$, which are related to the reflectivity and transmitivity such that $r=e^{i\phi}\sin(\theta)$ and $t=\cos(\theta)$. 
We have chosen the phase $\phi = \pi/2$ in Fig.~\ref{fig:phantm_macronode} such that a single application of the circuit at $\tau_i$ produces an output cat with the same orientation as the input cat, allowing repeated applications of PhANTM at $\tau_{i+1}$ and later timesteps. Note that there are two homodyne measurements applied in a unit round of PhANTM (one on $\nu_a$ and another on $\nu_{b}$) before the output mode is ready for the next round.
Also note that given the beamsplitters and phases involved in the circuit in Fig.~\ref{fig:phantm_macronode}, the bases used for either mode teleportation or deletion differ from the usual $p$ or $q$ bases for a canonical cluster state ~\cite{RaussendorfThesis}. 
The effective $C_Z$ gate in Fig.~\ref{fig:phantm_macronode} results from the application of two-mode squeezing, beamsplitters, and homodyne measurements as shown in Fig.~\ref{fig:cluster_ph}. The modes that undergo operation $\mathcal{O}$ in Fig. \ref{fig:chip_v2} have multiple photon subtractions followed by homodyne detection applied to them, while the rest of the modes in \textbf{IV} have $p$ basis homodyne measurement applied to them. Active displacements are needed in between repeated applications of PhANTM due to random homodyne measurement results. The value of the displacement is the output of the homodyne detector fed directly back to a standard displacement circuit with a local oscillator for the next temporal mode~\cite{EatonThesis2022}. The timescale of these active displacements must be shorter than the time delay between successive temporal modes in the cluster state.

\begin{figure}
    \centering
    \includegraphics[width = 0.45\textwidth]{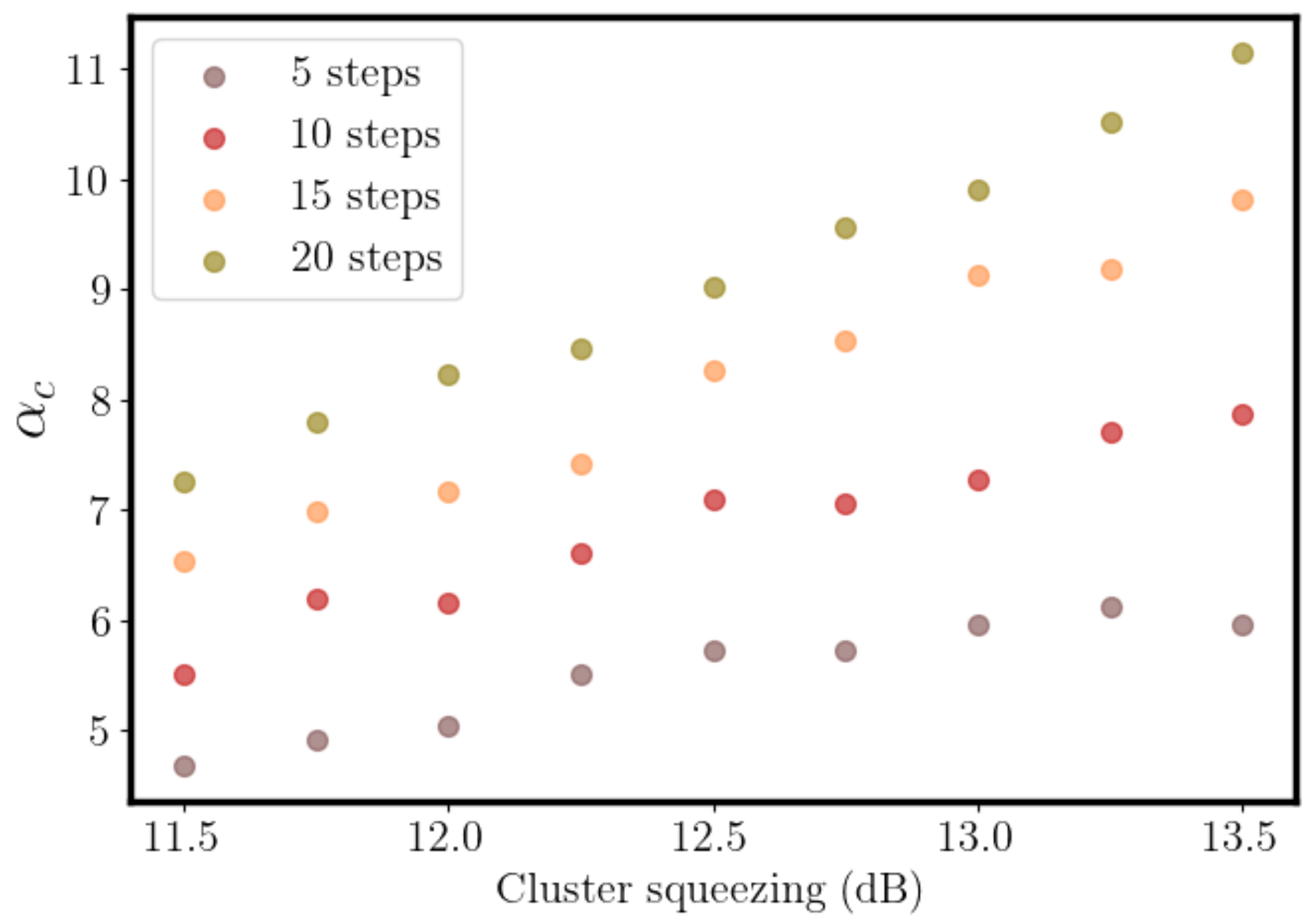}
    \caption{Mean $\alpha_c$ as a function of cluster squeezing. Each color corresponds to a given number of PhANTM steps. The means are estimated from 500 Monte Carlo iterations of PhANTM. Details of these simulation are given in Appendix. The plot shows the positive correlation of $\alpha_c$ with cluster squeezing and the total number of PhANTM steps.}
    \label{fig:mean_alpha_cor}
\end{figure}

Monte Carlo simulation of several rounds of PhANTM were conducted using the unit cell shown in Fig.~\ref{fig:phantm_macronode}, sampling the number of photons detected in each photon subtraction. The parameters of the cat states that we aim to optimize are their size and squeezing, both of which are stochastic. The larger these parameters, the higher the effective squeezing of the resulting GKP states, leading to lower error rates for QEC codes. 
To facilitate analysis and comparison across various conditions, a single parameter is defined: the amplitude of the cat state with zero squeezing ($\alpha_c$, also see Sec.~\ref{sec:GKP}).
Fig. \ref{fig:mean_alpha_cor} shows the mean of $\alpha_c$ as a function of cluster squeezing for different number of PhANTM rounds assuming $r_2=r_3$ (see also Sec. \ref{sec:MC simu} in Appendix). In these simulations, eight sequential photon subtraction attempts were performed at each time step. The beamsplitter angles for each photon subtraction are chosen such that no more than $10$ photons are detected per PNR detector. 
The subtraction probability goes down for later subtraction events, as continued subtraction attempts de-amplify the quantum signal.
We apply a reflectivity gradient (higher angle in later subtraction attempts) to ensure that the distribution of subtracted photons among the PNR detectors is roughly uniform (see Sec.~\ref{sec:app_pnr} in Appendix). 
Increasing the cluster squeezing enhances the potential number of photons that can be subtracted in a single subtraction event, resulting in cats with larger amplitudes. Furthermore, increasing the number of PhANTM steps raises the total number of photons subtracted across all cumulative PhANTM events, leading to larger amplitudes for cat states. 
The additional squeezing value also depends on the reflection and transmission coefficients of each beamsplitter utilized in the photon subtraction process. 

The success probability of each step of PhANTM, and the final cat size, can be tuned given the hardware constraints of the photon number resolution ceiling and the losses in the photonic circuit resulting from repeated teleportations in time, leading to many possible optimizations.

\section {Generation of GKP States}\label{sec:GKP}
There are two main approaches to creating GKP states, which we describe below for comparison with our approach.
In the first approach, single shot methods are devised to directly create GKP states. In this approach, Gaussian Boson Sampling (GBS)-based approaches to GKP state generation~\cite{tzitrin2020progress} produce probabilities of $2$\% to $7$\% with fidelities of around $0.65$ for GKP states with $10$ dB effective squeezing with $12$ dB of Gaussian squeezing. The probabilities are of O$(10^{-5}$) if the target fidelities are close to $1$. These probabilities are low for practical quantum computation. Bifurcation-based methods proposed creating $10$ dB GKP states with maximum photon number resolution of $40$ photons, Gaussian squeezing of up to $19.4$ dB with much higher success probabilities for very high fidelity GKP states~\cite{Takase2024}. The overall probabilities, while impressive, are still far from deterministic and resolving $40$ photons would require Transition Edge Sensors (TESs) working at cold cryogenic temperatures of $0.1$ K and low repetition rates of $100$~KHz~\cite{Gerrits2016}. 
Very high squeezing required also poses a challenge in physical implementation of the protocol. In another proposal, single-photon states are used in conjunction with GBS circuits to increase the fidelity to $0.95$ and success probability from $10$\% to $40$\%~\cite{Crescimanna2024}, not including the below-unity probability of generating single photons. In the second approach GKP states are generated from a set of cat states with a fixed $\alpha$ and parity by teleporting the cats through each other, also known as ``breeding''~\cite{Weigand2018}. In a practical implementation, deterministically creating photonic cat states with the same amplitude and parity is difficult. In short, GKP generation schemes either have low success probability, require high photon number resolution, use Fock states that are difficult to prepare with high fidelities and probabilities, or impose conditions on cat state parameters that are difficult to satisfy in a hardware implementation.  

We focus on creating GKP states above the FT threshold from breeding a supply of probabilistic cat states from PhANTM. 
Since PhANTM is a probabilistic process, fixed-$\alpha$ and fixed-parity cats are difficult to engineer. In addition, with a low probability the cat states produced from PhANTM have amplitudes that are too small for breeding. Finally, given a finite amount of squeezing in the cluster state, some cats from PhANTM are too large to be squeezed to the amplitude required for breeding. In the following, we present an adaptive breeding protocol that overcomes the probabilistic nature of PhANTM to create GKP states with a high effective squeezing.   

The pseudoalgorithm for adaptive breeding is shown in Algorithm~\ref{GKPalg}. The algorithm takes a set of cat states embedded in a cluster state. In order to breed GKP or sensor states with the correct spacing, the cat amplitudes must be adjusted using a squeezing operator.  
Once the amplitudes are adjusted, the cats are simply measured in the $p$ basis. Because the cats are embedded in a cluster state with effective $C_Z$ operations applied between the cluster modes, squeezing and breeding is implemented via measurements only~\cite{Eaton2022Phantm}.

\begin{algorithm}
\caption{GKP adaptive breeding algorithm} \label{GKPalg}
\raggedright
\textbf{Input} cluster squeezing $r_{cluster}$ \\
\textbf{Input} $M$ breeding steps \\
\textbf{Input} $2^M$ outputs from PhANTM, each with: unsqueezed cat amplitude $\alpha$, parity $P$, cat squeezing $r_{cat}$ \\
\textbf{Input} desired GKP state spacing $\xi$ \\
\textbf{Input} desired cat $\alpha_{b}$ in \cite{Weigand2018} $\alpha_{b} = \xi 2^{(M-3)/2}$ \\
\textbf{Input} lower limit of acceptable cat squeezing $r_{lb}$ \\
\textbf{Input} measurement bases $p$ and $q$

\begin{algorithmic}
\For{each PhANTM output}
    \If{$P$ is odd}
        \State displace cat to change phase by $\pi/2$
    \EndIf
    \State compute $s_{\alpha_b} = \ln(\alpha_b/\alpha)$ and $s_{2\alpha_b} = \ln(2\alpha_b/\alpha)$
    \If{$s_{\alpha_b} < r_{cluster}$ and $s_{\alpha_b} > r_{lb}$}
        \State squeeze cat with $s_{\alpha_b} - r_{cat}$
        \State teleport to next time step
    \ElsIf{$s_{2\alpha_b} < r_{cluster}$ and $s_{2\alpha_b} > r_{lb}$}
        \State squeeze cat with $s_{2\alpha_b} - r_{cat}$
        \State teleport to next time step
    \Else
        \State replace with momentum squeezed state 
    \EndIf
\EndFor

\State begin Breeding Procedure:
\For{each output pair produced above}
    \State select unmeasured modes on the cluster state on which to teleport output bred states
    \State breed and entangle outputs by measuring one mode of pair with $p$
    \EndFor
\For{each unmeasured mode remaining above}
    \State select unmeasured modes on the cluster state on which to teleport output bred states
    \State breed and entangle outputs by measuring one mode of pair with $p$
\EndFor
\State repeat Breeding Procedure until only one mode remains unmeasured 
\end{algorithmic}
\end{algorithm}

The adaptive breeding protocol takes several inputs: the number of desired breeding rounds, $M$; the squeezing present in the cluster state, $r_{cluster}$; a number of input cats which depends on the number of breeding rounds, $2^M$; the input cats' unsqueezed amplitudes $\alpha$, their present squeezing level after PhANTM, $r_{cat}$, and their parity $P$; the desired GKP or sensor state spacing in phase space, $\xi$; the required cat state amplitude for breeding, $\alpha_{b}$, which depends on the spacing as  $\alpha_{b} = \xi 2^{(M-3)/2}$ \cite{Weigand2018}; and the lower bound on cat squeezing that will determine its suitability for breeding, $r_{lb}$. All homodyne measurements are performed in the appropriate basis depending on whether an input cat is used in breeding or replaced with a squeezed vacuum.
\begin{figure*}[t!]
    \centering
    \includegraphics[width=1\linewidth]{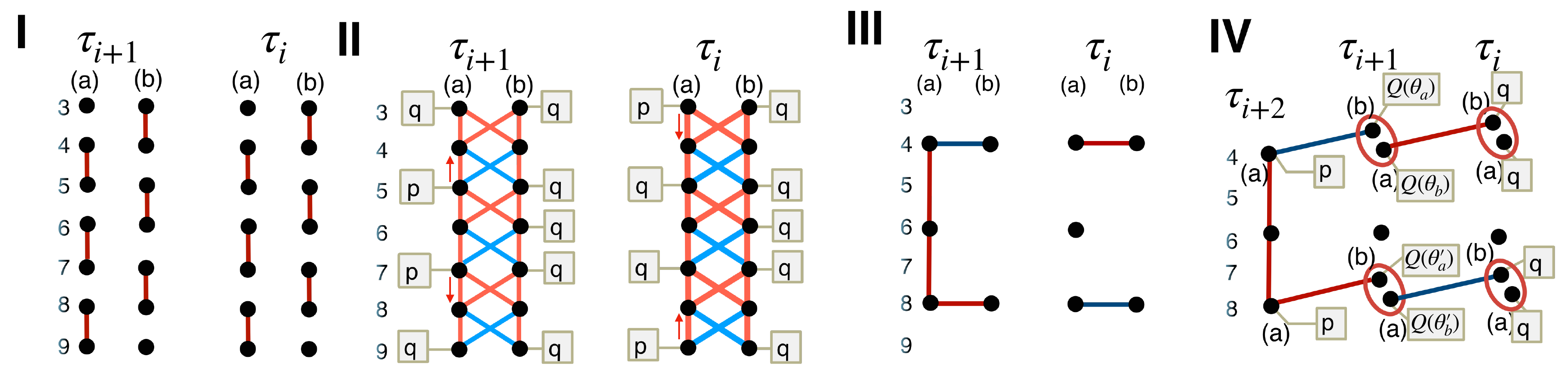}
    \caption{Cluster engineering for adaptive breeding. While the figure follows the same structure as in Fig. \ref{fig:cluster_ph}, a notable difference is the connection between dual rails of different frequency modes required here. Breeding is applied between states in modes $4_a$ and $8_a$ at the time step $\tau_{i+2}$ in \textbf{IV}. The output state of the breeding protocol is in mode $6_a$. Also at stage \textbf{IV} and prior to breeding, squeezing gates are applied at time step $\tau_{i+1}$ by setting angles of the homodyne detections. Note that the physical components on the chip create the cluster state in \textbf{I} and \textbf{II}, which is then engineered for different processes in \textbf{III} and \textbf{IV} by changing homodyne measurements. 
    }
    \label{fig:cluster_Breedin}
\end{figure*} 
As a first requirement for the adaptive breeding to proceed, the amplitude $\alpha$ of the cats output from PhANTM must be known. The output amplitude and squeezing, $r_{cat}$ depends on the photon subtraction vector obtained from PhANTM. 
The parity $P$ of each cat also depends on the total number of photons subtracted during PhANTM. As each cat can only breed with another cat of the same parity, small displacements can be used to adjust the odd cat parities to even ones. 
This displacement causes an effective, known displacement in the first breeding round output, which can be kept track of but does not have to be actively fixed until breeding is complete.
In general, the cat amplitudes are adjusted via squeezing operations of magnitude $s_{\alpha_b}$, which depends on the desired spacing $\xi$ and the number of breeding steps $M$.

The extra teleportation step in the breeding protocol allows us to introduce momentum squeezed states at $r_{cluster}$ squeezing in two scenarios. 
If the cat amplitude produced by PhANTM is too small to be used in GKP generation such that no amount squeezing smaller than $r_{cluster}$ can be used to adjust its amplitude to $\alpha_b$, it can be replaced by a momentum squeezed state before breeding. 
Adding a squeezed vacuum state in place of a cat increases the squeezing in one quadrature at the expense of the squeezing in the other quadrature. 
In the second scenario, arbitrarily large cat states can be produced in PhANTM that are too big to be squeezed to amplitude $\alpha_b$ needed in breeding because the squeezing required to do so would exceed $r_{cluster}$.
These cats can be squeezed to an even multiple of the required amplitude for the breeding protocol to proceed ($2\alpha_b$ is used in Algorithm~\ref{GKPalg}). 
If the squeezing of these large cats, $s_{2\alpha_b}$, is smaller than the desired lower bound, $r_{lb}$, then these cats can once again be replaced by momentum squeezed states (and likewise if the required squeezing would exceed $r_{cluster}$).
The squeezing of cats or cluster is the dominant term that determines the effective squeezing of the GKP states in one quadrature, while the number of breeding rounds is the dominant term for effective squeezing in the other direction.
Note that using a cat that is an even multiple of the required cat squeezing also has the effect of increasing the effective squeezing in one of the quadratures, giving us flexibility in trading off the squeezing in the GKP quadratures by combining mixtures of cats with different amplitudes and momentum squeezed states.

Fig.~\ref{fig:cluster_Breedin} illustrates the cluster engineering method required to apply adaptive breeding in Algorithm~\ref{GKPalg} to cat states generated on different dual rails of the cluster state (see also Appendix Sec.~\ref{sec:cluster_eng_squeezing_gate}). At stage \textbf{II}, the mode arrangement is modified compared with Fig.~\ref{fig:cluster_ph}. Specifically, prior to applying any breeding operation, distinct frequency modes need to be connected via an effective $C_Z$ interaction. This connection is shown in stage \textbf{III}. Breeding of cat states in mode $4$ and $8$ is applied in stage \textbf{IV} at the time step $\tau_{i+2}$. 
Prior to adaptive breeding, squeezing gates are employed to ensure that the size of each cat state matches $\alpha_b$ or an even multiple of $\alpha_b$ in the pseudoalgorithm above.
This operation requires one single time step and consumes two spatial modes of the same frequency as it is represented between time step $\tau_{i+1}$ and $\tau_{i+2}$ in Fig. \ref{fig:cluster_Breedin}. The squeezing parameter of the gate is set by the homodyne measurement angles $\theta_a$ and $\theta_b$ \cite{Alexander2014}. More details are discussed in Sec.~\ref{sec:cluster_eng_squeezing_gate} of the Appendix.  
Note that we can swap between sensor states and GKP qubits by changing the amplitudes to which the cats are squeezed. 

Notice that applying the squeezing operation or replacing cat states with momentum squeezed states does not require any active switching of the quantum states. Replacement of cat states is accomplished by setting $\theta_a=\pi/2$ and $\theta_b=0$ in homodyne measurements on the cluster (see Fig. \ref{fig:cluster_Breedin}). This causes the cluster mode in the next time step to be a momentum squeezed state, which is subsequently used for breeding, removing the need for active switching with an extra teleportation step.

To quantitatively model the adaptive breeding step, the outputs of the previously described PhANTM simulations are used as input to an adaptive breeding simulation. A Monte Carlo approach is used to obtain statistics about the sensor states obtained from adaptive breeding, with 1000 samples drawn from a set of input cats that were produced with an array of settings for starting cluster squeezing and the number of photon subtraction attempts. The simulation makes use of Qutip~\cite{JOHANSSON20131234} for density matrix calculations and Strawberry Fields~\cite{killoranStrawberryFieldsSoftware2019} for homodyne detection simulation.
The average sensor state squeezing along with their standard deviations are shown alongside FT threshold simulations in Sec.~\ref{thresh_sim}.

\section {Generation of Magic States}\label{sec:magic}
\begin{figure}
    \centering 
    \includegraphics[width = 0.45\textwidth]{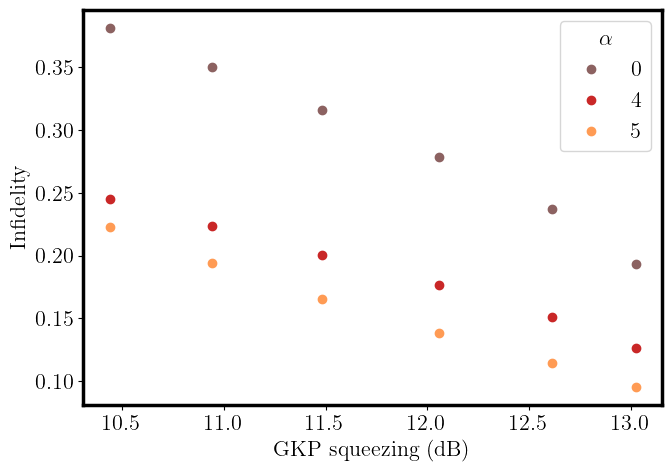}\\ \includegraphics[width = 0.45\textwidth]{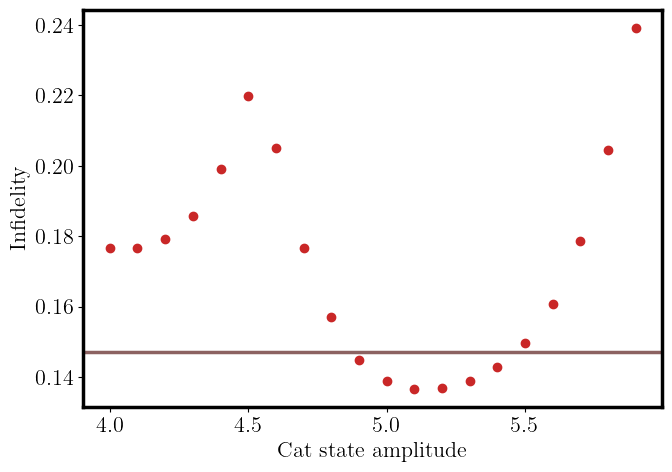}
    \caption{Top: magic state infidelity as a function of squeezing in the GKP states used as a QEC resource for three input cat amplitudes ($\alpha$), where $\alpha=0$ corresponds to the vacuum state. Bottom: magic state infidelity as a function of input cat amplitude and GKP squeezing of 13~dB. The horizontal line represents the distillation threshold for H-type states. }
    \label{fig:magicqec}
\end{figure}

Magic states are an important ingredient of universal FTQC as they allow implementation of non-Clifford operations while themselves remaining distillable with only Clifford gates~\cite{Bravyi2005}.  
A popular gate set which generates the Clifford group is \{$CNOT$, $H$, $S$\}, where $CNOT$ is the controlled NOT gate; $H$ is the Hadamard gate, and $S$ is the phase gate. 
The non-Clifford T gate usually completes the universal gate set~\cite{Bravyi2005}.
Provided with a source of physical magic states, one can distill a resource state that enables teleportation of non-Clifford operations into the QEC substrate at arbitrarily low error rates. 
Improving magic state quality by increasing the fidelity with respect to any of the $\ket{H}$ or $\ket{T}$ states reduces the physical resources required for a given logical error rate in the quantum computation.  
A promising method to produce a GKP magic state uses GKP error correction on the vacuum state~\cite{Baragiola2019}, implying that GKP qubits and encoded Clifford operations are sufficient for universal QC~\cite{Baragiola2019, calcluth2023vacuum}. However, simulations show that a reasonable success rate with existing methods requires the effective squeezing of GKP states to be extraordinarily high, approaching $20$ dB (see Fig.~\ref{fig:magicqec}), to create distillable magic states in one error correction round. 
In other architectures, producing magic states with practical success rates require the GKP states to have very high effective squeezing~\cite{Baragiola2019,Takase2024}, well above the FT threshold.  
In this section we provide a method to circumvent this issue and produce magic states with GKP resources that have effective squeezing levels near the FT threshold.

\begin{figure}
    \includegraphics[width=0.5\textwidth]{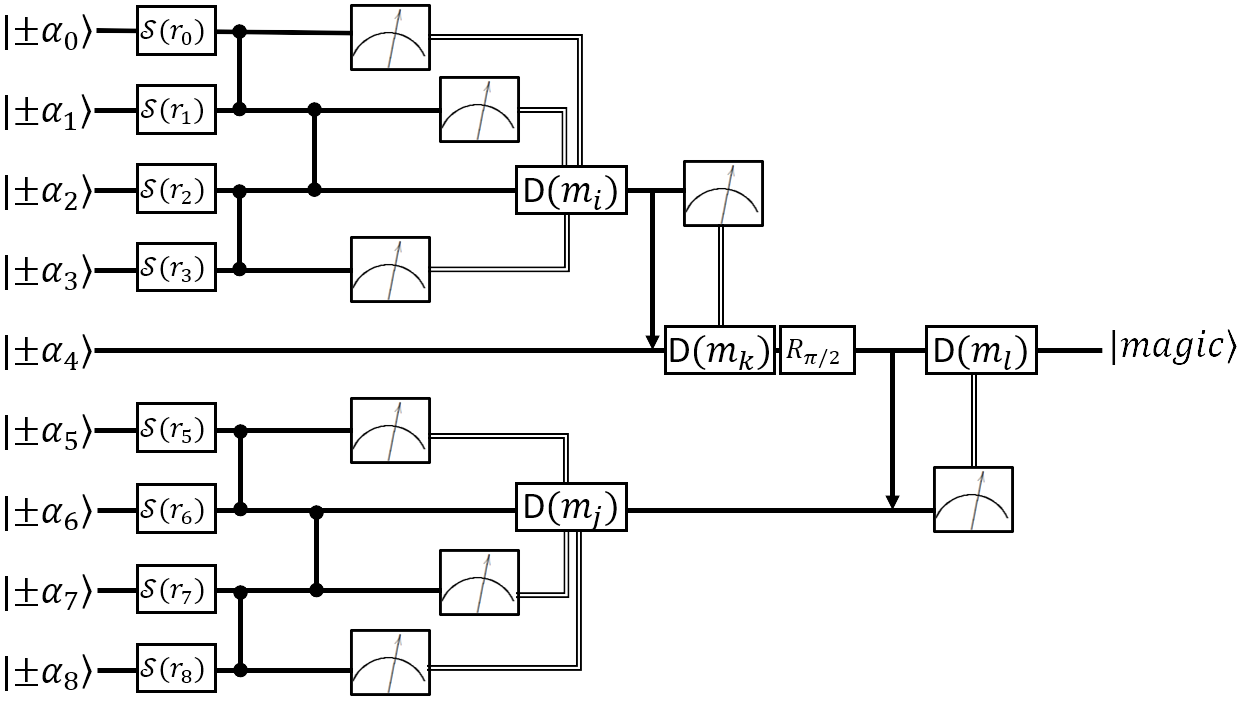}
    \caption{Magic state generation protocol. Double lines represent feed-forward operations. The gauge boxes represent $p$ basis measurements, except for the final measurement which is in the $q$ basis. The unsqueezed input cats labeled $\ket{\pm\alpha_i}$ can also be replaced with prior cascaded breeding steps in order to increase the quality of the bred GKP states before applying CV-QEC.}
    \label{fig:cat_magic}
\end{figure}

The method used to create magic states with significant probability using much lower squeezed GKP states in the present architecture is shown in Fig.~\ref{fig:cat_magic}. 
We replace the input vacuum with a cat state ($\ket{\pm\alpha_4}$ in Fig.~\ref{fig:cat_magic}) on which to apply QEC with GKP states. This substitution opens up the possibility to optimize the input cat amplitude, as well as all operations in the GKP breeding process prior to CV-QEC. 
We also replace the $C_Z$ operations during CV-QEC with beamsplitters, which allows us to use separate photonic chip components to produce GKP and cat states. 

This substitution requires a single squeezing operation to be performed on one of the GKP resource states, which is easily commuted to the state preparation stage before breeding the GKP state. 
A simple implementation of the protocol is shown in Fig.~\ref{fig:cat_magic}, resulting in a magic state generation circuit using nine cat states. This contains the equivalent of two GKP breeding rounds for each of the two GKP resource states that are consumed; the circuit with 3 GKP breeding rounds would contain 17 input cats. 
The cat input $\ket{\pm\alpha_4}$ breeds successively with the output of two cat trees, each of which produces high quality GKP states.

The distillation infidelity threshold for H-type states is approximately 14.7\%. The plots in Fig.~\ref{fig:magicqec} show a deterministic simulation of CV-QEC on input vacuum or cat states with post-selection on quadrature measurement outcomes of zero. Note that this post-selection, done to ease simulations, will slightly underestimate the output squeezing of the bred GKP states \cite{Weigand2018}. We see that when performing CV-QEC on the vacuum mode (purple data points) one obtains infidelities above the distillation threshold (of 14.7\%) when using GKP states with squeezing of 13dB or less as the resource state for CV-QEC. On the other hand, with effective squeezing greater than 12dB and cat states with amplitude of $\sim5.1$, magic states with infidelities below the distillation threshold can be obtained.

\begin{figure}
    \centering
    \includegraphics[width=1\linewidth]{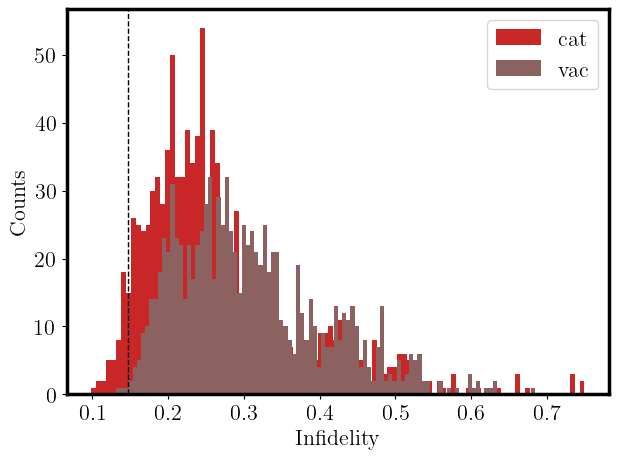}
    \caption{Histograms of fidelities obtained from Monte Carlo simulation of magic state generation for CV QEC on vacuum modes (brown) or cat states (red).}
    \label{fig:maghisto}
\end{figure}

In practice, the cat states produced by PhANTM have random amplitudes, and they must be squeezed in order to obtain the correct spacing for GKP breeding before magic state generation. This means that our GKP states have a random squeezing distribution. Further, as Fig.~\ref{fig:magicqec} shows, the optimal input cat amplitude for the error-corrected mode is between 4.9 and 5.5. 
Because our cats have random amplitudes and squeezing levels, only a few will hit this required target, although PhANTM can be engineered to emit cat states with mean squeezing levels and amplitudes centered on the requirements.
Monte Carlo simulations were used to determine the success rate of producing a magic state when random cats from PhANTM were used to provide the resource states, both for GKP breeding and the cat state upon which CV QEC was performed. The simulations use Qutip~\cite{JOHANSSON20131234} for density matrix calculations and Strawberry Fields~\cite{killoranStrawberryFieldsSoftware2019} for homodyne detection simulation.
Fig.~\ref{fig:maghisto} shows histograms of the infidelities obtained from each magic state generation attempt, where the fidelity is measured with respect to an ideal $H$-type magic state. 
The cat states were produced using Monte Carlo PhANTM simulations with a cluster state squeezing of 13.5~dB, 20 PhANTM steps, and 8 photon-subtraction attempts made before homodyne detection and teleportation to the next PhANTM time step.
When using vacuum states as the error-corrected mode, we see a success rate of 0.003 after 1000 iterations (brown counts in the histogram). 
On the other hand, when using cat states as the error-corrected mode, we see a success rate of 0.048 after 1000 magic state generation attempts, an improvement of more than factor of 10, which reduces resources required to produce a magic state by a factor 10 relative to using vacuum alone. 
This simulation specifically targeted $H$-type states, which enable teleportation of $T$ gates within the architecture.
These success rates mean that magic factories will not require the majority of qubit resources in surface code substrates using lattice surgery~\cite{litinski2019magic}.

\section {Encoding Logical Qubits in a Surface Code}\label{logical}
GKP encoding can protect against small displacement errors, but larger displacements ($>\frac{\pi}{2}$) will result in an error. Depending on the quadrature, this can be either a Pauli X error ($q$ quadrature) or a Pauli Z error ($p$ quadrature). These logical qubit errors require additional qubit-level error-correction. Many error correcting codes have been studied in the context of GKP qubits \cite{Bourassa2021blueprintscalable,stafford2023biased,raveendran2022finite,zhang2021quantum}. We will focus on topological error correction based on a family of codes known as surface codes, which have been shown to have high thresholds ($\approx 1\%$ under circuit level noise \cite{wang2011surface} or $\approx 10 \mathrm{dB}$ with GKP states \cite{Bourassa2021blueprintscalable}) and well-studied logical gates \cite{litinski2019game}.

\begin{figure*}[t]
    \centering
    \includegraphics[width = 0.9\textwidth]{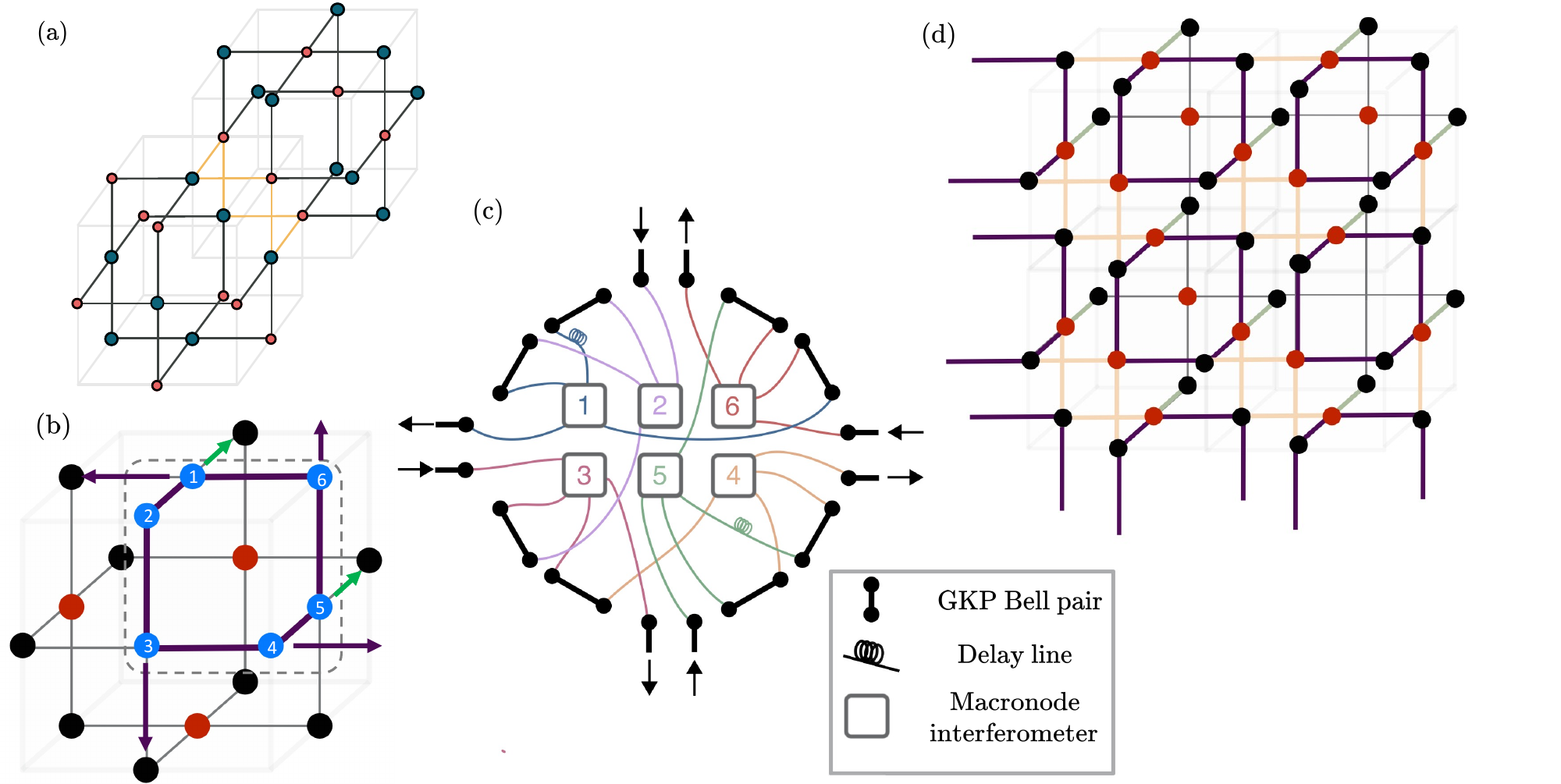}
\caption[unit cell]{(a) RHG unit cells. Primal unit cell shown with blue qubits on faces and red qubits on edges. Products of $X$-basis measurement on face qubits are a stabilizer of the cluster state. Dual unit cell shown shifted by half a unit in all directions. Now red qubits are on faces and blue qubits are on edges. By measuring all primal and dual stabilizers, Pauli errors on all qubits can be corrected. (b) Six macronode tilable unit cells shown relative to the RHG unit cell. Purple arrows show where GKP nodes are sent and green arrows dictate delayed modes. (c) Macronode layout within the unit cell. 12 GKP Bell pairs are created per unit cell with four modes being sent to neighboring unit cells and four being received. Temporal connections between macronodes are created using delay lines. The colored links represent fiber optic connections between separate optical chips for the GKP generation and macronode interferometer circuits. (d) Example of the tileability of the unit cell into a larger RHG lattice. Orange links denote Bell pairs shared between unit cells and green edges indicate temporal links. Bell pairs can be generated by interfering sensor states output from the GKP generation algorithm on a beamsplitter. }
    \label{fig:unit_cell}
\end{figure*}

As mentioned earlier, photonic CVQC is well-suited to MBQC. Circuit model codes known as Calderbank-Steane-Short(CSS) codes, where X and Z errors are detected independently, can be translated into a cluster state for fault-tolerant MBQC via a process known as foliation \cite{bolt2016foliated}, which loosely maps the data qubits in the circuit model code to linear cluster states.
Surface codes are examples of CSS codes, and the resulting foliated 3D cluster state is often called the Raussendorf-Harrington-Goyal (RHG) lattice after the authors who first introduced it \cite{Raussendorf2007topological}. The unit cell of this cluster state is shown in Fig. \ref{fig:unit_cell}. The product of $X$-basis measurements (equivalently $p$-basis homodyne measurements for GKP qubits) on all face qubits is a \textit{stabilizer} of the cluster state. 
In the error free case, measuring stabilizers will result in +1 eigenvalue and errors will result in $-1$ measurement results for any stabilizer that they anti-commute with.
A second unit cell can be defined by shifting the base unit cell by 1/2 in all directions. Here, the edge qubits of the initial unit cell become face qubits of the second unit cell. These two lattices are known as primal and dual lattices, respectively. By measuring the stabilizers of both lattices, errors on all qubits can be detected. Logical operators in foliated cluster states are formed by correlation surfaces that map logical operators from the underlying code from one end of the cluster state to the other. To implement logical X and Z operations, correlation surfaces of the RHG lattice are given by 2D sheets of qubits which span the lattice and intersect primal and dual unit cells, respectively. The code distance is the shortest chain of errors that can result in a logical error, which is equivalent to the shortest error chain that can span the RHG lattice. The threshold theorem states that provided the physical error rate of the underlying qubits is below some threshold, we can arbitrarily suppress the logical error rate by increasing the code distance \cite{knill1998resilient}.

The popular Clifford + T gate set maps naturally to the surface code. Clifford gates and teleportation based T gates can be implemented by lattice surgery~\cite{litinski2019game}. This approach defines logical qubits as patches, which are subsections of the overall surface code lattice, with logical operators identified on the edges. In MBQC, this corresponds to volumes of the RHG lattice with correlation surfaces on the boundaries. Patches can be merged by measuring stabilizer checks between edges of the patches, thus performing a joint logical measurement of the two patches. Clifford gates and the T gate can be implemented with joint measurements including ancilla qubits or magic states. In fact, in Ref.~\cite{litinski2019game}, it was shown how all Clifford gates can be commuted through the T gates and measurements, resulting in circuits that just consist of mulit-qubit joint measurements including a magic state. Lattice surgery operations can also be mapped to changing measurement bases (homodyne angles) in fault-tolerant MBQC \cite{herr2018lattice}.

This gate set is well studied and optimized, however our architecture can be adapted to other gate sets \cite{litinski2019game,gidney2019flexible} and other error correcting codes \cite{walshe2024linear}. In this paper we aim to quantify Gaussian squeezing resources needed to demonstrate FT thresholds with the RHG lattice serving as the example code. We leave optimization of the error correcting code and gate set for future work. 

\subsection {Macronode GKP lattice construction}\label{sec:macronode}

With GKP qubits there exists more than one method for constructing the RHG lattice in Fig \ref{fig:unit_cell}(a). 

We can construct the lattice directly, where the nodes are GKP $\ket{+}$ states and the edges are CV $C_Z$ gates, which can be implemented deterministically with linear optics and squeezing. This approach has two main draw backs. Firstly, while it has been demonstrated previously~\cite{Yoshikawa2008}, in-line squeezing is generally difficult to realize experimentally.
Second, ideal $C_Z$ gates are not physical, requiring 
infinite squeezing. Approximating the gate with finite squeezing will add noise to the GKP qubits.
Alternatively, any lattice can be constructed with passive interferometers starting with GKP Bell pairs, where the number of physical modes per macronode is the valency of the underlying \emph{canonical} graph state and the GKP Bell pairs are distributed to match the entanglement in the canonical graph state. 
Relating measurements on the physical modes with effective measurements on the underlying lattice requires a list of corrections known as a \textit{dictionary protocol} \cite{Alexander2014}, which we describe in Appendix~\ref{app:macronode}. Note, that in the following we use the terms GKP state and sensor state interchangeably, always meaning either a sensor state or one half of a GKP Bell pair formed by interfering sensor states.

We wish to design a modular construction of the required lattice. To this end, we identify a unit cell consisting of 24 sensor states or 12 GKP Bell pairs that can be tiled to construct a macronode RHG lattice of arbitrary size. From this unit cell, four GKP nodes are sent to neighboring unit cells, and correspondingly four are received. Two of the modes are delayed to form temporal links. This is illustrated in Fig.~\ref{fig:unit_cell}. As the primal and dual layers are created in a single time-step, we avoid the requirement for adaptive optics that have often been used to account for the different entanglement structure in primal and dual layers \cite{Bourassa2021blueprintscalable,larsen2021fault}. To teleport non-Clifford operations onto logical qubits within the surface code, physical magic states are injected into the RHG lattice~\cite{litinski2019magic,herr2018lattice}. The physical magic state is interfered with a vacuum state at a beamsplitter and is used in place of one GKP Bell pair in the RHG macronode. 

We note here that due to the inherent randomness in the measurements, there are several feed forward corrections that need to be accounted for. Some of them, such as the corrections due to the macronode construction and the random outcomes on the effective Pauli measurements on the RHG lattice, do not need to be physically implemented and can be tracked in software. While certain operations do require physically implementing the feedforward, such as changing measurement bases due to magic state injection, changing the measurement basis corresponds to changing the homodyne phase for GKP based architectures. In this way, the required adaptivity at the QEC level does not require any fast modulators on the qubit modes.

\section {FTQC Threshold Simulation}\label{thresh_sim}
Several CV FTQC architectures based on GKP qubits and the surface code exist in the literature and have been characterised by their GKP effective squeezing threshold. Refs.~\cite{Fukui2018} and \cite{Bourassa2021blueprintscalable} used constructions based on $C_Z$ gates and found thresholds of $10$ dB and $10.5$ dB respectively, however these ignored the additional noise from finite squeezing in physical gates. In Ref.~\cite{larsen2021fault} found a threshold of $12.7$ dB while accounting for finite squeezing. Ref.~\cite{tzitrin2021fault} proposed a macronode-based architecture which did away with the need for $C_Z$ gates and found a threshold of $10.1$ dB. Refs.~\cite{Bourassa2021blueprintscalable,larsen2021fault,tzitrin2021fault} all require fast reconfigurable optics either in the computation or in the preparation of GKP states with a sufficiently high success probability.

In the following section, we analyze the fault-tolerant performance of our architecture. In order to allow comparison with other architectures, we first use a phenomenological error model used widely in the literature \cite{Fukui2018,fukui2021all,larsen2021fault}, which models realistic GKP states as ideal GKP states acted on by a Gaussian displacement channel.  Formally, this is represented as \cite{tzitrin2021fault}
\small
\begin{equation}
    \mathcal{E}\left(\rho,\Delta\right) = \frac{1}{\sqrt{\pi}\Delta^2}\int \mathrm{d}\mathbf{\alpha} \exp\left(-\frac{1}{\Delta^2}\left|\mathbf{\alpha} \right|^2\right)\hat{D}\left(\mathbf{\alpha} \right)\rho\hat{D}^\dagger\left(\mathbf{\alpha}\right)
\end{equation}\label{eq:gauss_disp_channel} 
\normalsize
Which has the effect of replacing the delta function teeth of an ideal GKP state with Gaussian distributions with a standard deviation, $\Delta$,where $\Delta$ describes the same broadening of the GKP teeth as $\Delta_D$ described in Sec.~\ref{sec:CVtools} \cite{Bourassa2021blueprintscalable}. We can link this to the GKP effective squeezing in dB as $S_\mathrm{dB} = - \log_{10}\left(\Delta^2\right)$~\cite{Bourassa2021blueprintscalable}. 
Despite its simplicity, this error model can describe many common noise sources such as finite squeezing and photon loss \cite{tzitrin2021fault,Bourassa2021blueprintscalable,larsen2021fault,noh2020fault}. 
We then move on to quantify the fault-tolerance performance in terms of the input Gaussian resource for PhANTM - the squeezing level of the multi-mode cluster state. The input for these simulations are distributions of GKP squeezings obtained from simulations of the proposed GKP generation pipeline.

In both cases, to determine the logical error rate we track how the GKP effective squeezing propagates through the optical circuit used to generate the macronode lattice and use the propagated noise variances to simulate noisy homodyne measurements. We apply the corresponding post-processing to reduce the physical macronode measurements to the effective noisy measurements on the canonical lattice, as shown in Appendix~\ref{app:macronode}. With these noisy measurements we determine the probability of a qubit level error on each node of the canonical lattice, see Appendix~\ref{app:inner_decoder} for more detail.  These probabilities are used to sample a set of errors and generate the corresponding syndrome of the RHG lattice. This syndrome is then decoded using the pyMatching implementation of the minimum weight perfect matching decoder~\cite{higgott2023sparse}. This simulation method is described in more detail in Appendix~\ref{app:sim_method} and is similar to the method used in Ref \cite{Bourassa2021blueprintscalable}, adapted to the macronode approach in Ref.~\cite{tzitrin2021fault} and for our specific GKP generation method. 

\subsection{Phenomenological error model}
Here, we present results which tailor the phenomenological noise model to the PhANTM and adaptive breeding approach to generate GKP states. As described in Sec.~\ref{sec:GKP}, generating GKP states from PHANTM and breeding results in GKP states that are inherently asymmetric due to the different physical processes that determine the squeezing in the GKP quadratures. Note that this is equivalent to grid states with asymmetric spacing in the two quadratures \cite{stafford2023biased} which has also been studied in \cite{walshe2024linear}.
This feature was incorporated into our simulation method by defining $p$ and $q$ noise variances separately for each GKP state. While we can also set each GKP state independently, for simplicity we assume that that the $p$ and $q$ noise are identical for each GKP state. To determine the 2D threshold curve for unbalanced quadrature noise we employ the method used in Ref.~\cite{Bartolucci2021}. By expressing the $p$ and $q$ noise in terms of a dummy variable, $\Delta_p = c_p \times x$, $\Delta_q = c_q \times x$, we fix the ratio of $p$ and $q$ noise via $c_p/c_q$ and use $x$ to find the threshold for a given ratio. 
\begin{figure*}[t!]
    \centering
    \includegraphics[width = 0.9\textwidth]{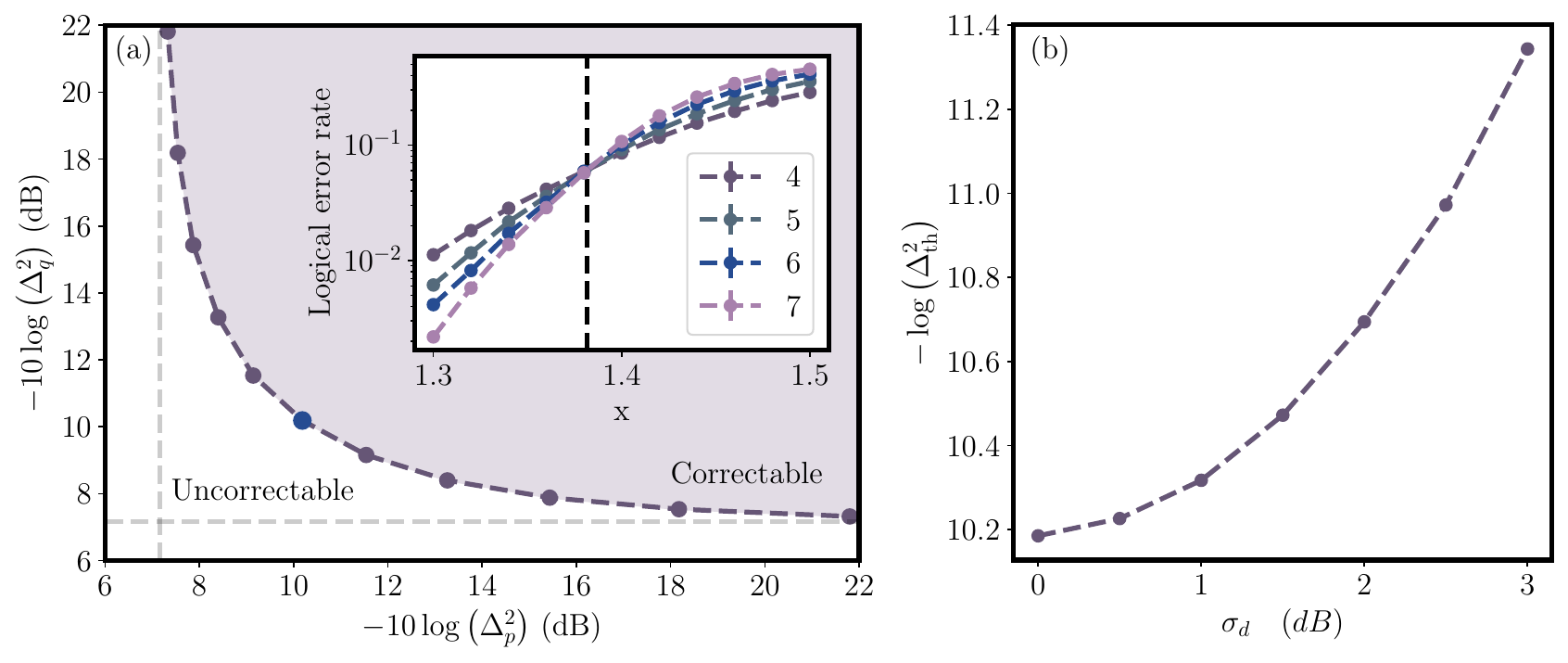}
\caption[Independent noise]{(a) 2D threshold plot for independent noise in the GKP quadratures, in terms of decibel squeezing, $S_\mathrm{dB} = -10\log\left(\Delta^2\right)$. Combinations of squeezing values above the line are correctable and those below are not. Dotted lines show the marginals with infinite squeezing in one of the quadratures. The inset shows an example threshold plot for a single point, shown in blue, in the main panel. Each point is averaged $10^5$ times. (b) Plot showing the dependence of the squeezing threshold on the distribution of GKP effective squeezing.  On each Monte Carlo run, first the GKP effective squeezing for each mode from a normal distribution with standard deviation $\sigma_d$ before the noisy homodyne values are generated from the sampled squeezing value.}
    \label{fig:biased_noise}
\end{figure*} 
Fig.~\ref{fig:biased_noise} shows the resultant surface, where we have shaded the correctable region. The surface is symmetric with marginal squeezing thresholds of $7.2 \mathrm{dB}$. With balanced squeezing in each quadrature we recover a threshold of $10.2 \mathrm{dB}$, which is inline with similar error models \cite{tzitrin2021fault}.

A second adaption stems from the fact that the size of the cat states from PhANTM depend on the number of photons subtracted and so are inherently probabilistic. This results in GKP states that do not have a single squeezing value for each quadrature, but rather are described by a distribution of effective squeezing. To include this in our simulation framework we employ a two-stage sampling procedure where for each mode and for each Monte Carlo iteration we sample first the squeezing variance for each mode from a Gaussian distribution with standard deviation given by $\sigma_d$ and then we sample the noisy homodyne value. Fig.~\ref{fig:biased_noise}(b) shows how the squeezing threshold depends on $\sigma_d$. 

\subsection{End-to-end simulations}

As in many similar architectures \cite{larsen2021fault,Fukui2018,chamberland2019fault,Bourassa2021blueprintscalable,tzitrin2021fault}, the previous section abstracted the thresholds in terms of GKP effective squeezing. While this allows a comparison against different architectures, it does not give an estimate of the physical requirements for our approach. A key resource for our architecture is the squeezing level of the multi-mode Gaussian cluster state. In this section we aim to determine the threshold in terms of this resource. To this end we perform simulations of PhANTM and adaptive breeding and then use the resultant GKP states as the inputs to the logical-level simulations. We fix the PhANTM and breeding setup to use eight PNR detectors per frequency mode with gradient beamsplitters, see Sec.~\ref{sec:phantm}. We chose three breeding rounds and employ a threshold for cat squeezing of $r_{lb}=0.5$ ($4.34 \mathrm{dB})$, below which a momentum squeezed state from the cluster state is swapped in, as described in Sec.~\ref{sec:GKP}. We then sweep the cluster squeezing and simulate the logical error rate for several code distances, repeating this simulation for a range of PhANTM steps.

\begin{figure*}[t!]
    \centering
    \includegraphics[width = 0.9\textwidth]{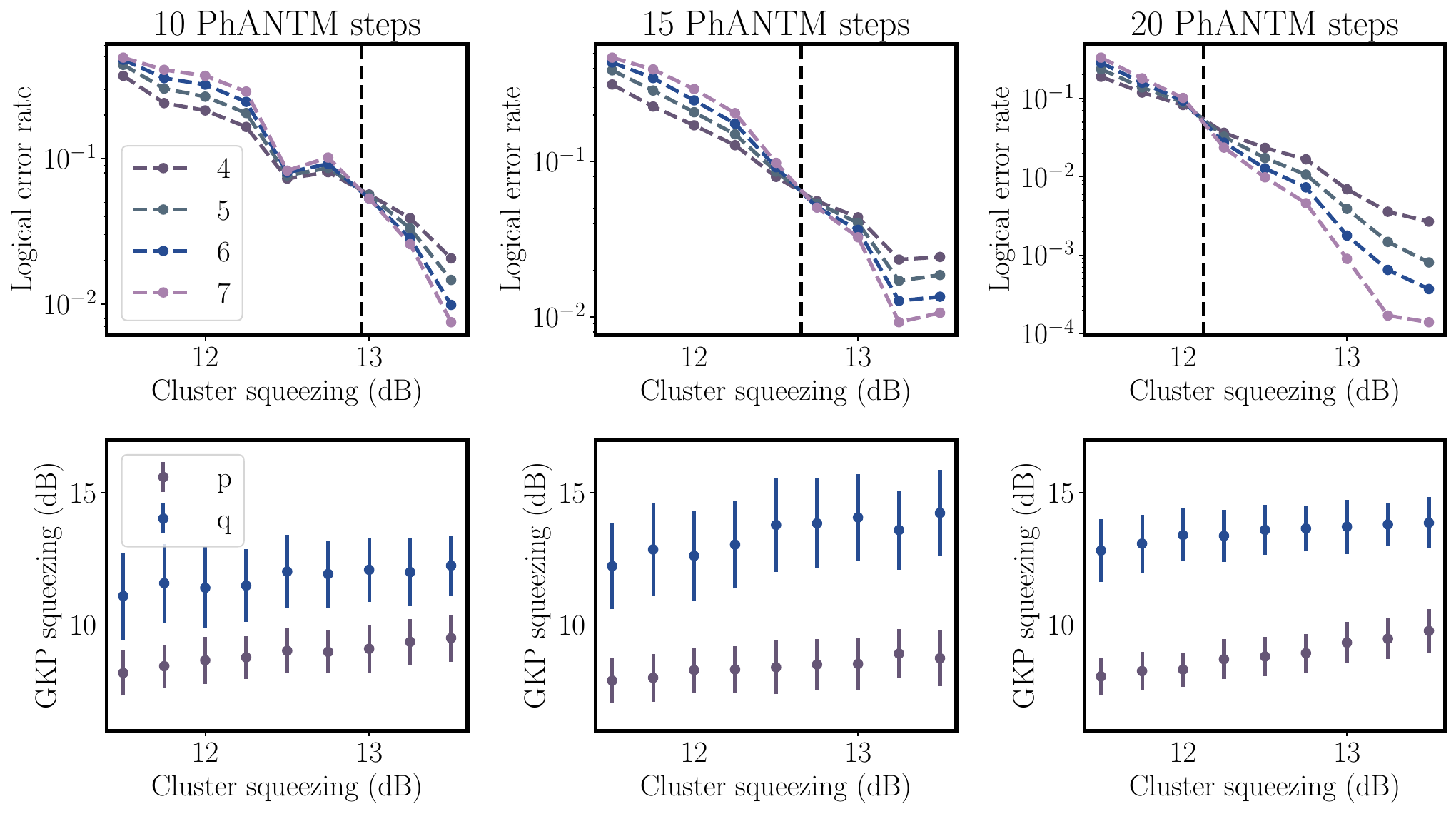}
    \caption[cluster threshold]{Fault-tolerant threshold for cluster squeezing for a range of PhANTM step numbers. Bottom panels show the corresponding mean and standard deviations of GKP effective squeezing distributions input to the logical-level simulation. These results used $10^3$ PhANTM iterations, $10^3$ adaptive breeding iterations and $10^5$ logical error rate trials.}
    \label{fig:distr_sqz}
\end{figure*} 

We find thresholds for cluster squeezing needed for PhANTM and teleportation-based squeezing gate of $12.1$, $12.7$, and $13$ dB for $20$, $15$ and $10$  PhANTM steps, respectively. We see that the required cluster squeezing increases as the number of PhANTM steps, and therefore the mean cat state amplitude $\alpha$, decreases. The lower panels in Fig.~\ref{fig:distr_sqz} show the mean and standard deviation of the GKP effective squeezing distributions output from the breeding process. We also compare the adaptive breeding protocol, used in Fig.~\ref{fig:distr_sqz}, to standard breeding, where cat states must be squeezed exactly to the GKP spacing. With standard breeding, no threshold was found below $13.5$ dB, indicating that adaptive breeding can be used to reduce the threshold by \emph{at least} $1.5$ dB and possibly more.
While our simulation takes into account all generated GKP states, not just those above threshold, we can quantify the proportion of GKP states that fall above threshold, allowing us to not only compare the Gaussian squeezing resources required but also the generation probability with other GKP generation protocols. To this end, we determine the likelihood that a GKP state drawn randomly from the generated distribution has a squeezing level above the threshold, that is the combination of squeezing in each quadrature puts it in the 'correctable' region of Fig.~\ref{fig:biased_noise}. This likelihood can then be taken as the success probability of our protocol. Some examples are shown in table~\ref{tab:GKPsuc}.

\begin{table}[b]
\begin{center}
\begin{tabular}{|c||c c|} 
 \hline
 Cluster squeezing (dB) & \multicolumn{2}{c|}{PhANTM steps}  \\ [0.5ex] 
 & 10 & 20 \\ 
 \hline
 12.0 & 0.26 & 0.45 \\
 \hline
 13.0 & 0.56 & 0.91 \\ [0.5ex] 
 \hline
\end{tabular}
\caption{Success rates for which adaptive breeding produces sensor states with squeezing levels above the FT threshold for different levels of cluster squeezing and PhANTM steps.}\label{tab:GKPsuc}
\end{center}
\end{table}

While direct simulations of the required Gaussian squeezing thresholds for other GKP generation methods are missing from the literature, we can compare to several approaches that benchmark themselves by generating $10$ dB GKP states, which are at the threshold for several fault-tolerant architectures \cite{tzitrin2021fault,fukui2023high,Fukui2018}. The highest success probability for single-shot GKP generation protocols are found in Ref.~\cite{Takase2024}; however they require input $18-20$ dB of Gaussian squeezing, and photon number resolution up to 40 photons. On the other hand, lower input squeezing resources can be used as in Ref.~\cite{tzitrin2020progress}, requiring $12$ dB, and Ref.~\cite{Winnel2024} is as low as $6 \mathrm{dB}$. However, in the former, success probabilities are on the order of $10^{-5}$ and the latter requires the generation of large, $n=10$, Fock states which are experimentally challenging .

\section{Discussion}

We have presented an architecture for CV FTQC that uses the time-frequency-space degrees of freedom to create GKP states above the FT threshold with high probabilities, creates distillable magic states with a supply of realistically squeezed GKP states and cat states, and can be implemented on low-loss photonic chips in a modular architecture. This work proposes a static state generation process that produces a distribution of GKP effective squeezing, affecting the FT threshold only modestly compared to idealized GKP states with a set squeezing target, as shown in Fig.~\ref{fig:biased_noise}. The results presented in the paper require Gaussian squeezing levels of $12$ dB to $13$ dB. Low photon number resolution is needed for cat state generation after which all steps require homodyne detections and linear optical CV operations. The architecture can be implemented without on-chip switches or quantum memories, resulting in a fully-passive architecture. 

The experimental hardware needed for the architecture is ambitious but feasible. The squeezing level needed is below the state-of-the-art levels achieved in free-space~\cite{Vahlbruch2016}. In recent years, the squeezing measured from on-chip sources has steadily improved, reaching $8.3$ dB~\cite{kashiwazaki2023over}, with efforts underway to reach levels seen in free-space. The other major resource needed in our approach is photon number resolving detection. While TES devices offer resolutions in excess of $100$ photons~\cite{eaton2022photons} and have traditionally been used in CVQC, SMSPDs~\cite{kong2024} with low data acquisition requirements operating at higher temperatures have been shown to detect up to $10$ photons, sufficient for our architecture. 
Simultaneously, chip-integrated room-temperature PNR technologies capable of detecting up to $10$ photons at MHz scales are also in development~\cite{nehra2020photon}. Finally, while losses in photonic circuit components need to be minimized for implementing our architecture on photonic chips, Silicon nitride platforms currently provide some of the lowest losses available for passive components~\cite{psiquantum2025manufacturable} at levels that look promising for FTQC.

Multiple improvements, optimizations, and validations of the presented architecture remain to be explored. Higher probability methods for cat state generation via optimizing the squeezing and photon number resolution are possible. Adaptive breeding, used in both sensor state and magic state generation protocols, can be optimized with advanced adaptation algorithms. In the case of magic states, targeting a higher success probability while simultaneously reducing the quality of GKP states needed for magic generation looks possible with further development of the CV-QEC process with non-Gaussian states. The squeezing requirements of the architecture could also be lowered by moving to cluster states with a lower valency \cite{paesani2023high} and employing more sophisticated GKP decoders, such as the ones taking into account correlations introduced by linear optics \cite{walshe2024linear}. 

While the effect of losses in PhANTM has been investigated earlier~\cite{Eaton2022Phantm}, appendix \ref{sec:loss} provides simulations and a discussion of the effect of losses on the adaptive breeding and QEC protocols. We briefly note that PhANTM is more sensitive to propagation and homodyne losses than to photon number resolution losses, and shows significant reduction in the Wigner negativity of cat states for losses beyond 1\%. As for breeding, the effective squeezing of GKP states from breeding is reduced by less than $1$ dB with 3\% homodyne loss. Similarly, the GKP effective squeezing needed to be at $10.2$ dB threshold for fault tolerance is increased to about $11.6$ dB with a 5\% loss. This demonstrates that losses are less important for adaptive breeding and QEC than for cat generation through PhANTM.  Estimating the loss and noise targets especially with further design improvements would require more comprehensive error models and simulations, including the effects of Gaussian noise on effective squeezing and optical loss on Wigner negativity at each step of the architecture. An end-to-end simulation incorporating losses is left for future work.

While many improvements could be made and the performance under losses are yet to be fully quantified, we believe that the modular and passive architecture for photonic CV FTQC presented in the paper will constitute a foundational design for fault tolerant photonic quantum computers in the future.

\section{Acknowledgements}
Part of the QEC simulation code was funded under DARPA Contract HR00112390051. The views,
opinions, and/or findings expressed are those of the author(s) and should not be interpreted as representing the
official views or policies of the Department of Defense or the U.S. Government. We acknowledge insightful conversations with QC82 co-founders Olivier Pfister, Xu Yi, Joe Campbell, and Andreas Beling on theoretical and experimental CVQC, squeezing sources, homodyne detection, photon number resolution, and many other topics. We are thankful to Leandre Brunel for sharing some of his work on GKP lattice construction methods.

\bibliography{bibfile}

\section{Appendix}

\subsection{Additional terminology} \label{appterm}
Wavefunction for the GKP qubits in the $q$ basis can be defined as $\ket{\mu}_G\sim\sum_{n=-\infty}^\infty\ketsub{(2n+\mu)\sqrt{\pi}}{q}$
where $\mu\in\{0,1\}$, and superpositions can be defined as usual: $\ket{\pm}_G=\tfrac{1}{\sqrt{2}}(\ket{0}_G\pm\ket{1}_G)$~\cite{EatonThesis2022}. 
Any superposition of these states has support in phase space only on integer multiples of $\sqrt{\pi}$, and each state is periodic with period $2\sqrt{\pi}$. Because of this, small shifts in either position or momentum can be corrected. 
Note that a squeezed cat state has the form of the zeroth-order GKP state, in light of which it is possible to convert between Cat states and GKP states using breeding protocols.  
The sensor state, which we use to construct logical GKP qubits within a surface code, is defined as

\begin{align}
    \ket{S} \sim \sum_{n=-\infty}^\infty\ketsub{n\sqrt{2\pi}}{q}.
\end{align}

There are two families of logical magic states, the T-type and the H-type magic states~\cite{garcia_alvarez2020}, defined as
\begin{align}
\label{eq:magic_states}
    \ket{T}& = \cos\frac{\theta}{2}\ket{0}_G + e^{i\pi/4}\sin\frac{\theta}{2}\ket{1}_G \\
    \ket{H}& = \frac{1}{\sqrt{2}}(\ket{0}_G+e^{i\pi/2}\ket{1}_G),
\end{align}
where $\theta = \arccos(1/\sqrt{3})$. 
There are $8$ T-type and $12$ H-type magic states obtainable by applying Clifford operations on $\ket{T}$ and $\ket{H}$. 

\subsection{Demonstration of PhANTM on dual-rail wire}

Eq. \ref{eq:circuit_phantm} represents the PhANTM circuit corresponding to the operation in stage \textbf{IV} of Fig. \ref{fig:cluster_ph}. However, this three-mode circuit is not optimal for numerical simulations. In what follows, we demonstrate how the circuit can be reduced to a two-mode operation efficient for simulation purposes. We have
\begin{equation}
    \begin{quantikz}
        \ket{\Psi_{in}} & & & \arrow[d] &  \gate{O} &  \ket{m_1}_{p1} \\
        \ket{0}  & \gate{\mathcal{S}(r_2)}  &  \ctrl{1}  & & &\ket{m_2}_{p2} \\
        \ket{0}  & \gate{\mathcal{S}(r_3)} & \control{} & & & \ket{\Psi_{out}}
    \end{quantikz} 
\label{eq:circuit_phantm}
\end{equation}
where arrow depicts a balanced beamsplitter with $\pi/2$ phase. Using the definition of the beamsplitter, we can write the relation
\begin{equation}
    B_{\frac{\pi}{2}, (12)}C_{Z_{(23)}}B_{\frac{\pi}{2}, (12)}^\dagger = \frac{1}{\sqrt{2}}\mathrm{e}^{\frac{i}{\sqrt{2}}(q_2+p_1)q3},
\end{equation}
where $B_{\frac{\pi}{2}, (12)}$ and $C_{Z_{(23)}}$ stand respectively for beamsplitter between mode 1 and 2 with a $\pi/2$ phase and $C_Z$ gate between modes 2 and 3. From this relation we deduce the commutation relation between the operators as
\begin{equation}
\begin{aligned}
    &B_{\frac{\pi}{2}, (12)}C_{Z_{(23)}} = \\ & \qquad C_{Z_{(23)}}(1/\sqrt{2})R_1(-\frac{\pi}{2})C_{Z_{(13)}}(1/\sqrt{2})R_1^\dagger(-\frac{\pi}{2})B_{\frac{\pi}{2}},
\end{aligned}
\end{equation}
where $R$ ie the rotation operator defined as $R(\theta) = e^{i\theta a^\dagger a}$.
This relation can be used in the PhANTM circuit above to get
\\
\begin{equation}
    \begin{adjustbox}{width = 0.5\textwidth}
    \begin{quantikz}
        \ket{\Psi_{in}} & & \arrow[d] & \gate{R(\pi/2)} & \ctrl[wire style={"g"}]{2} &  \gate{R(-\pi/2)} & \gate{O} &  \ket{m_1}_{p1} \\
        \ket{0}  & \gate{\mathcal{S}(r_2)} &   &    & & \ctrl[wire style={"g"}]{1} &\ket{m_2}_{p2} \\
        \ket{0}  & \gate{\mathcal{S}(r_3)}&  & &  \control{} &\control{} & \ket{\Psi_{out}}
    \end{quantikz}
    \end{adjustbox}
\end{equation}
with $g=1/\sqrt{2}$.
The weighting of the $C_Z$ gate can utilize $\mathrm{e}^{igq_2q_3} = \mathcal{S}^\dagger(\mathrm{ln}(g))\mathrm{e}^{iq_2q_3}\mathcal{S}^\dagger(\mathrm{ln}(g))$ \cite{Eaton2022Phantm}. Subsequently we introduce the squeezing parameter $r'$ such that $\mathcal{S}(r')=\mathcal{S}(\mathrm{ln}(g))\mathcal{S}(r_3)$. Furthermore since the $C_Z$ gates commute, the circuit becomes
\begin{equation}
\begin{adjustbox}{width = 0.5\textwidth}
    \begin{quantikz}
        \ket{\Psi_{in}} & & \arrow[d] & & \gate{R(\pi/2)} &   \ctrl[wire style={"g"}]{2}&  \gate{R(-\pi/2)} & \gate{O} &  \ket{m_1}_{p1} \\
        \ket{0}  & \gate{\mathcal{S}(r_2)} & & \ctrl{1}   & \ket{m_2}_{p2} \\
        \ket{0}  & \gate{\mathcal{S}(r')}&  & \control{} & \gate{\mathcal{S}^\dagger(\mathrm{ln}(g))} &\control{} & \ket{\Psi_{out}}
    \end{quantikz}
    \end{adjustbox}
\end{equation}
On the third mode, we can use the expression
\begin{equation}
    \mathcal{S}(r')\ket{0} = G \ket{0}_{p3},
\end{equation}
where $G$ is defined as
\begin{equation}\label{eqn:G}
    G=\pi^{1/4}\sqrt{\frac{2}{s}}\mathrm{e}^{-q^2/2s^2}
\end{equation}
with $s=\mathrm{e}^{r'}$. Then $\ket{0}$ can be replaced by the p momentum squeezed state.
Knowing that $G$ can commute with $C_Z$, the circuit becomes

\begin{equation}
\begin{adjustbox}{width = 0.5\textwidth}
    \begin{quantikz}
        \ket{\Psi_{in}} & & \arrow[d] & & \gate{R(\pi/2)} & &  \ctrl[wire style={"g"}]{2}&  \gate{R(-\pi/2)} & \gate{O} &  \ket{m_1}_{p1} \\
        \ket{0}  & \gate{\mathcal{S}(r_2)} & & \ctrl{1}   & \ket{m_2}_{p2} \\
        \ket{0}_{p3}  & &  & \control{} & \gate{G} & \gate{\mathcal{S}^\dagger(\mathrm{ln}(g))} &\control{} & \ket{\Psi_{out}}
    \end{quantikz}
       \end{adjustbox}
\end{equation}

We recognize the teleportation circuit between mode 2 and 3. Subsequently, we obtain the final circuit presented in Fig. \ref{fig:phantm_macronode}.
\begin{widetext}
\begin{equation}
    \begin{quantikz}
        \ket{\Psi_{in}} &  &\arrow[d] & & & & \gate{R(\pi/2)}  &  \ctrl[wire style={"g"}]{1}&  \gate{R(-\pi/2)} & \gate{O} &  \ket{m_1}_{p1} \\
        \ket{0}  & \gate{\mathcal{S}(r_2)}& & \gate{Z^\dagger(m_2)} & \gate{R(\pi/2)} & \gate{G} & \gate{\mathcal{S}^\dagger(\mathrm{ln}(g))}  & \control{}&\ket{m_2}_{p2} \\
    \end{quantikz}
    \label{eq:phantm_2modes}
\end{equation}
\end{widetext}
From this circuit we can identify operator $\zeta$ as
\begin{equation}
    \zeta = Z^\dagger(m_2)R(\pi/2)G \mathcal{S}^\dagger(\mathrm{ln}(g)).
\end{equation}
The circuit in Eq. \ref{eq:phantm_2modes}, is the one we use to perform Monte Carlo simulation of PhANTM. $G$ operator has a second order effect and is not included to speed up run times with large Fock dimensions. That's because we take cluster squeezing higher than $11.5$ dB for which $2s^2\approx15>>1$ in Eq.~\ref{eqn:G}. 

\subsection{Monte Carlo simulation of PhANTM}
\label{sec:MC simu}
\begin{figure*}
    \centering
    \includegraphics[width = 1\textwidth]{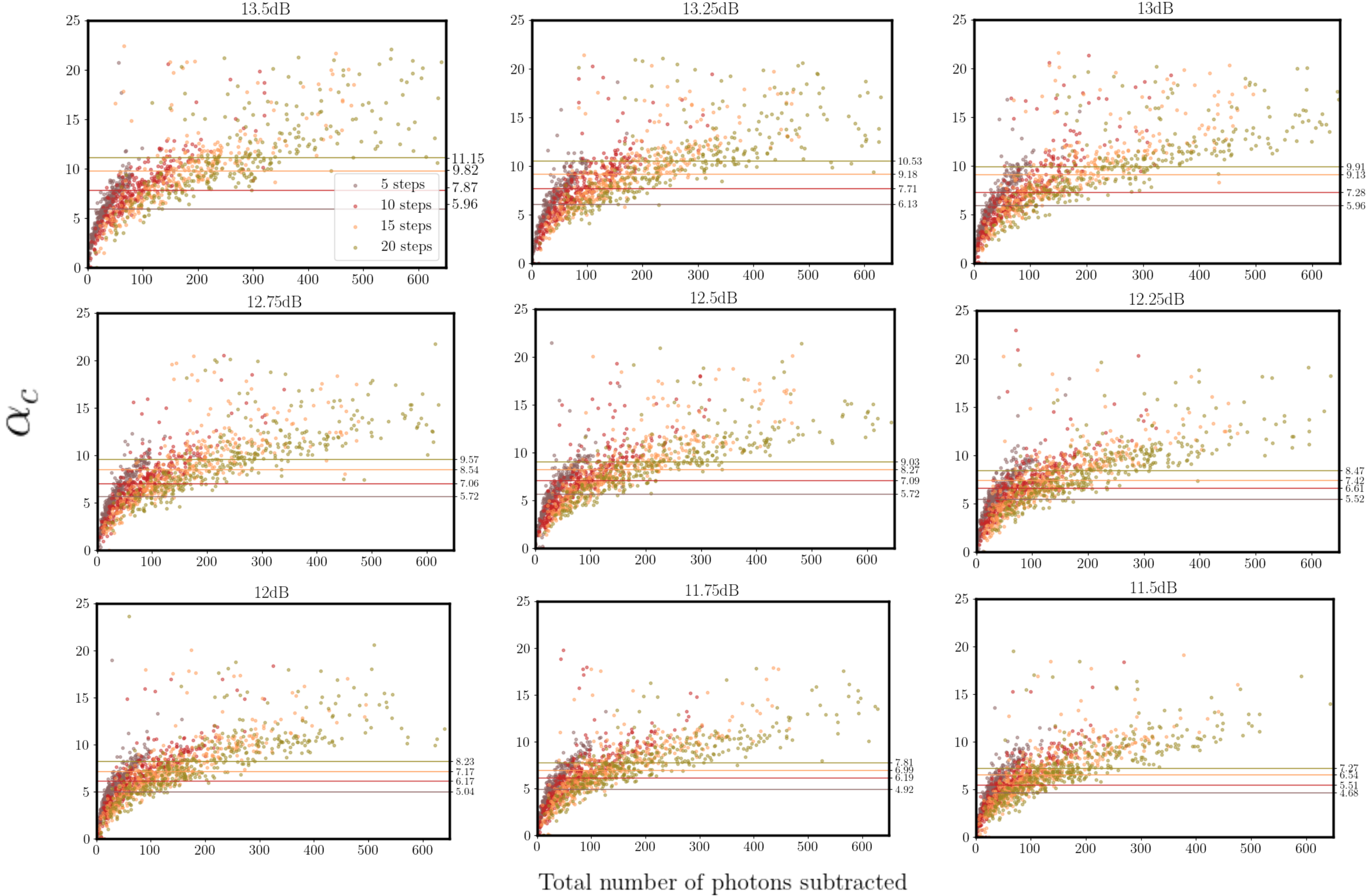}
    \caption{Results of PhANTM simulations (500 iterations for each dataset). Corrected alpha ($\alpha_c$) as a function of the total number of photons subtracted for different cluster squeezing and different number PhANTM steps. Horizontal line shows mean of $\alpha_c$ for each number of PhANTM step. Mean values are reported on the right axis of each plots}
    \label{fig:alpha_cor_fulldata}
\end{figure*}

The Monte Carlo methods used for PhANTM simulations make use of Qutip~\cite{JOHANSSON20131234} for density matrix calculations and Strawberry Fields~\cite{killoranStrawberryFieldsSoftware2019} for homodyne detection simulation.
Fig. \ref{fig:alpha_cor_fulldata} shows the results of the PhANTM simulation for various levels of cluster squeezing and $5, 10, 15$, and $20$ PhANTM steps.  
The $C_Z$ gate weight is set to 1.  
The number of photons subtracted at each subtraction event is determined stochastically, based on the density matrix of the state that is about to undergo photons subtraction~\cite{Eaton2022Phantm}. The stochastic nature of the photon subtraction results in a random cat state size and squeezing after the cumulative steps of PhANTM. 

For the simulations, a Fock dimension of 60 is used and 1000 trials are run.  
Additionally, to enhance computational efficiency, we post-select homodyne measurements at 0, thereby avoiding the needs for further operations to correct  the displacements induced at each PhANTM step. Effect of non zero homodyne measurement results on the weight unbalance of the cat states are discussed in \cite{Eaton2022Phantm}.

Each output state is fitted with a cat state wavefunction to determine the size and squeezing  as well as the parity of the cat state. To isolate dimensional issues in simulations, only states with a fidelity greater than $95\%$ are retained, and these are represented in Fig. \ref{fig:alpha_cor_fulldata}. These cat states are subsequently utilized for the adaptive breeding stage.

\subsection{Photon number resolution}\label{sec:app_pnr}
\begin{figure}
    \centering
    \includegraphics[width = 0.45\textwidth]{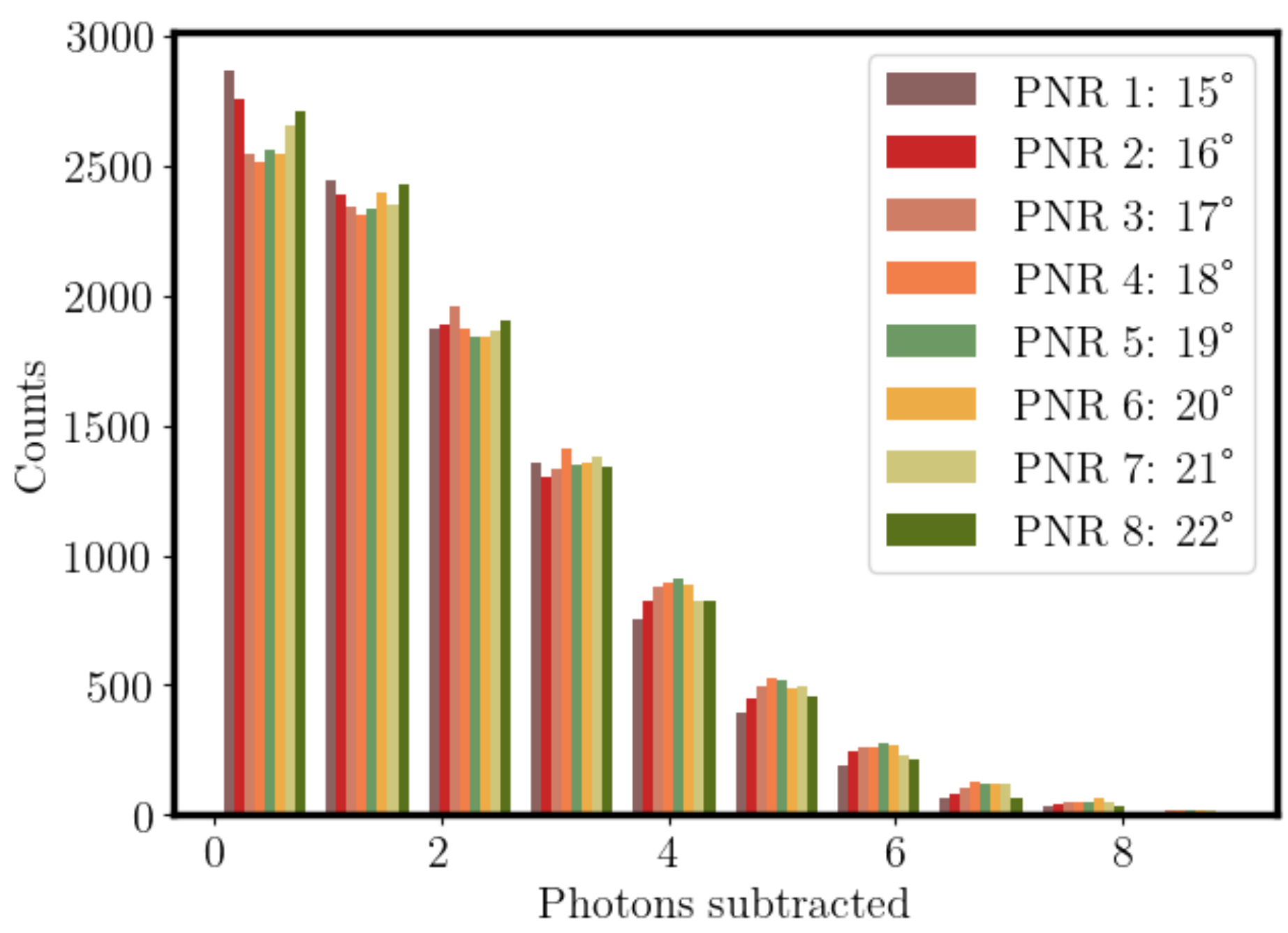}
    \caption{Histogram of photons detected for each of the 8 PNR detectors in the Cat generation process through 20 steps of PhANTM (500 iterations). Data are taken from Monte Carlo simulation of PhANTM with 13.5dB of cluster squeezing. Angle of the beamsplitter for each subtraction is given in the legend.}
    \label{fig:PNR_hist}
\end{figure}

Simulation results show that only up to 8 photon subtraction events are required to make high quality cat and GKP states. Fig. \ref{fig:PNR_hist}, shows a histogram of the results of photon subtraction measurements assuming eight PNR detectors in PhANTM Monte Carlo simulations.
The histograms appear largely uniform due to the gradient applied within the beamsplitter angle evolved in subtractions. CVQC schemes have traditionally needed photon resolutions of 20 to 50 photons to generate GKP states for FTQC with a success probability of a few percent (see Sec.~\ref{sec:GKP}). 

\begin{figure*}
\centering
{\includegraphics[width = 0.8\textwidth]{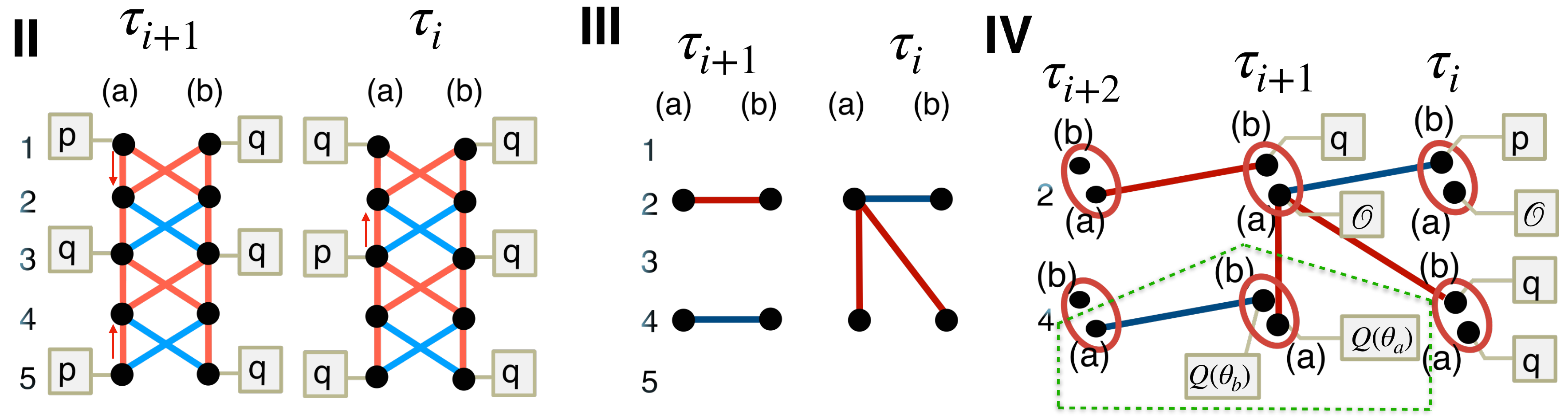}}
\caption{Cluster engineering for squeezing gate application. Stage \textbf{I} is not shown to avoid repetition with Fig. \ref{fig:cluster_ph}. In \textbf{IV}, is illustrated the application of the squeezing gate in mode 4 within the time step $\tau_{i+1}$ in the green dash box. The squeezing parameter is determined by the angles $\theta_a$ and $\theta_b$ of the homodyne detections. In mode 2, is performed a single PhANTM step, then the state is transferred to mode 4 before applying the squeezing gate. }
    \label{fig: Sgate_app}
\end{figure*} 

\subsection{Squeezing gate after PhANTM in dual rail cluster} \label{sec:cluster_eng_squeezing_gate}
In Fig. \ref{fig: Sgate_app}, it is shown that the squeezing gate is applied on a neighbouring mode of the dual rail where is applied PhANTM. Both operations can't be applied within the same dual rail without utilizing a switch since for PhANTM, a photon subtraction is required. At stage \textbf{II} and time step $\tau_i$, mode $3_a$ is measured in the $p$ basis to create then a $C_Z$ connection between mode 2 and 4. In \textbf{IV}, by applying $\mathcal{O}$ in mode 2 at the $\tau_{i+1}$ time step, a homodyne measurement is performed in the $p$ basis. Then the quantum state is teleported in mode 4 where corrected displacement can be applied. By choosing correctly local oscillator phase in homodyne detection at the $\tau_{i+1}$ time step, the squeezing gate can be applied from $\tau_{i+1}$ to $\tau_{i+2}$ in mode $4$. Derivation of the squeezing gate can be found in \cite{Alexander2014, walshe2020continuous} in the case of a beamsplitter with a phase $\phi=0$. In that case the squeezing gate, $\mathcal{S}$, can be written as
\begin{equation}
    \mathcal{S} = N_m C_m V_i
    \label{eq:Sgate}
\end{equation}
where $N_m$ and $C_m$ are respectively noise and displacement operators. $V_i$ can be expressed as:
\begin{equation}
    V_i= R(\theta_{i,+})S(\tan(\theta_{i,-})R(\theta_{i,+}))
\end{equation}
with 
\begin{equation}
    \theta_{i,\pm} = \frac{\theta_{i,a}\pm \theta_{i,b}}{2}
\end{equation}
where $\theta_{i,a}$ and $\theta_{i,b}$ are angles of homodyne detection. By using a beamsplitter with a phase of $\phi=\pi/2$, the formula \ref{eq:Sgate} remains valid, with the distinction that the angle $\theta_{i,a}$ and the input state are both rotated by $\pi/2$.

\subsection{Macronode RHG lattice construction}\label{app:macronode}

Here, we describe some of the specifics of the reduction from the macronode version to the RHG lattice to the canonical cluster state. A full derivation is given in ref \cite{tzitrin2021fault} so we will give only a brief introduction here. The protocol is illustrated in Fig. \ref{fig:dict_protocol}. 
A balanced four-mode splitter is used to distribute the entanglement. To recover the measurements on the canonical RHG lattice we assign one of the four physical nodes as the central mode, which we label as mode 0. This can be measured in either the $p$ or $q$ basis, depending on which basis we would wish to measure the RHG lattice node in. The remaining three auxiliary modes are always measured in the $q$ basis. To map from the physical measurement on the central and auxiliary modes to the equivalent canonical measurement result we require an appropriate dictionary protocol.
The dictionary protocol is derived uses the fact that a beam-splitter can be expressed as two $C_X$ gates and two inline squeezing gates. $C_X$ gates with control modes directly before a measurement can be replaced by a $q$ displacement on the target mode, dependent on the measured mode. The inline squeezing operations can be commuted through any remaining $C_X$ gates and displacements to the end of the circuit. The squeezing gates on any auxiliary modes are then absorbed into the $q$ basis measurement. This results in a single squeezing gate on the central mode, which re-scales the measurement results by 2 or 0.5 depending on the measurement basis, and an $X$ correction ($q$-basis displacement) that depends on the $q$ basis measurement results of the auxiliary modes. The remaining $C_X$ gates are commuted backwards past the GKP Bell pairs which introduces a new set of $C_Z$ gates which mirror those of the original lattice, and a set of $Z$ corrections ($p$-basis displacements) that depend on the measurement results from the neighboring macronodes. This results in the same entanglement structure as the canonical lattice, up to a squeezing operation and $X$ and $Z$ corrections on the central mode.

\subsection{GKP binning}\label{app:inner_decoder}

In this section we describe the process of mapping CV displacement errors to qubit-level errors. This is referred to as the "inner decoder" in Refs.~\cite{Bourassa2021blueprintscalable,tzitrin2021fault} as it is the process of decoding the GKP code.

We employ the simple binning process where each homodyne value is rounded to the nearest multiple of $\sqrt{\pi}$. Odd and even multiples define the two logical qubit states. The error probability can be determined by determining how much of the Gaussian distribution of each bin overlaps with the neighboring bins \cite{Fukui2018}. However, this approach loses some information that can be gained from the position of the homodyne value relative to its bin. For example, a homodyne value found in the center of a bin is much less likely to have come from the neighboring bin than one towards the bin edge. We can quantify the error probability for a specific homodyne value, $z$, and a GKP tooth width $\Delta$ as \cite{Fukui2018,Bourassa2021blueprintscalable}.

\begin{equation}
p_\mathrm{err}\left(z,\Delta\right) = \frac{\sum_{n \in \mathbb{Z}} \exp{\left[-\left(z - (2n + 1)/\sqrt{\pi}\right)^2/\Delta^2\right]}}{\sum_{n \in \mathbb{Z}} \exp{\left[-\left(z - n/\sqrt{\pi}\right)^2/\Delta^2\right]}}.
\end{equation} \label{eq:perr}
\begin{figure*}[t!]
    \centering
    \includegraphics[width = \textwidth]{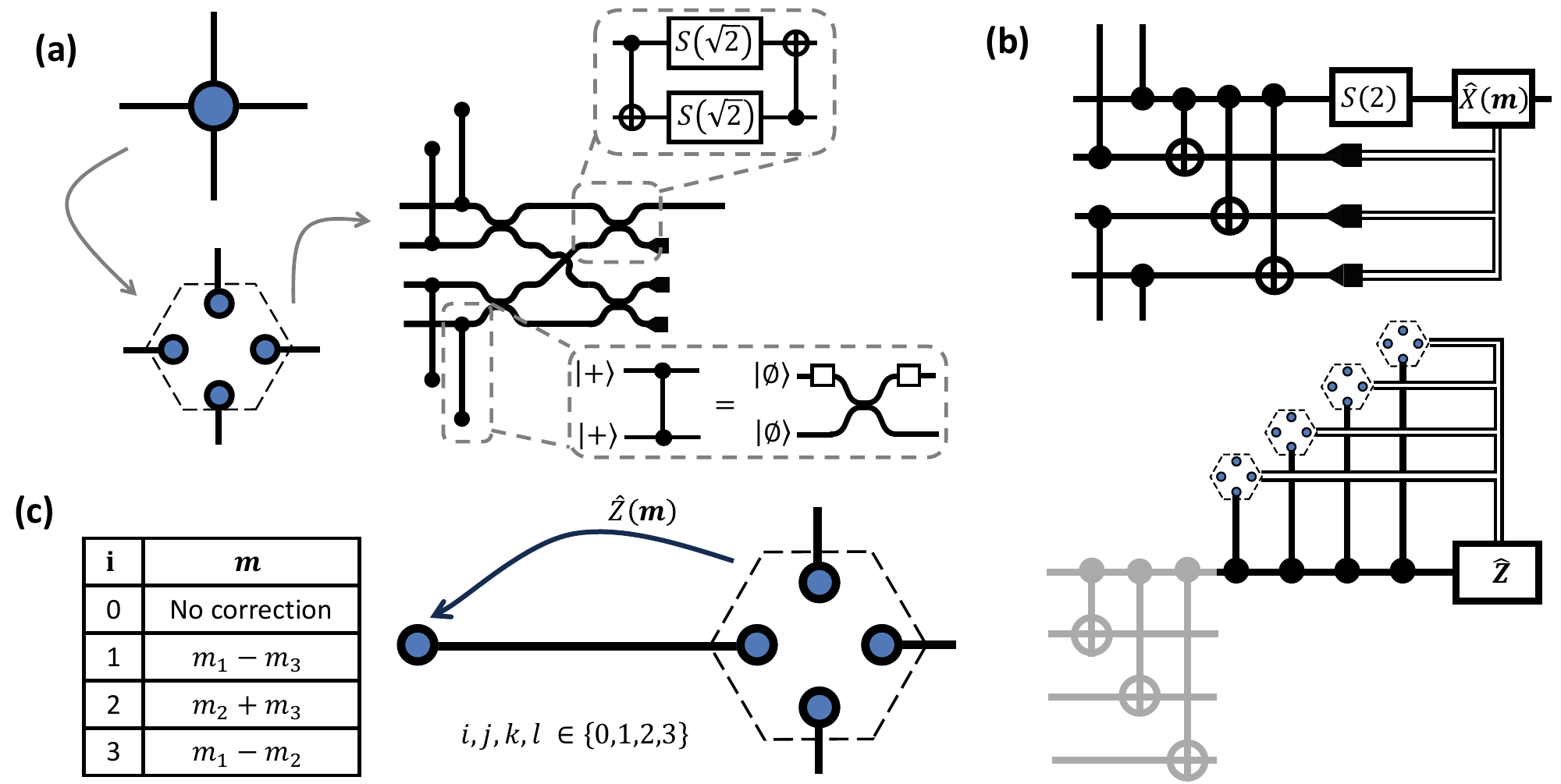}
\caption[Dictionary protocol]{ Illustration of the dictionary protocol for the macronode RHG lattice. a) Each node in the RHG lattice is replaced by a four node macronode. Bell pairs can be created by sensor states interfered on a passive beamsplitter (shown in the lower call-out box). A balanced four-mode splitter is used to generate the long range entanglement. Note, that while we have drawn a circuit with $C_Z$ for the Bell pair generation, the physical circuit only consists of beamsplitters and phase rotations (shown with white squares.) The dictionary protocol can be derived by noting that the beamsplitters can be represented as $C_X$ and squeezing gates (shown in the upper call-out box). By commuting these gates through each other we can arrive at the same entanglement structure as the canonical lattice with a squeezing correction and an $X$ and $Z$ corrections, corresponding to a $q$ and $p$ basis displacements, respectively. b) Effective circuit after the $X$ correction has been derived. The $X$ correction depends only measurements from auxiliary modes within the macronode. $\mathbf{m} = \frac{1}{2}\left(m_2 + m_3 - m_4\right)$ where $m_i$ is the $q$-basis measurement result from the $i^\mathrm{th}$ auxiliary mode. c) Effective circuit after the $X$ and $Z$ corrections have been derived, equivalent to the canonical RHG lattice. Commuting the three $C_X$ gates from (b) through the $C_Z$ gates adds extra $C_Z$ gates with support on the central mode. $C_Z$ gates with support on auxillary modes can be replaced with measurement based $Z$ corrections. Once commuted through the remaining $C_Z$ gates, the resulting $Z$ corrections on each central mode will now depend on measurement results from neighboring macronodes. The exact form of the correction depends on which mode the in the neighbor had support on the original $C_Z$ link. We show these corrections in the table. $i$ corresponds to mode index in the neighboring macronode. Mode 0 is the central mode. }
    \label{fig:dict_protocol}
\end{figure*} 

We are interested in the error rate on the effective measurements on the canonical lattice. That is after the physical measurements have been post processed. As we have GKP states on every physical mode in each macronode, it is sufficient to determine the probability that the central mode measurement and each correction is incorrectly binned \cite{tzitrin2021fault}. This is because, to first order, the total error probability of the macronode is given by the sum of these individual probabilities.

As well as the linear optics which can alter the GKP noise, the measurement result post processing also has an effect on the effective distribution that the combined corrections are sampled from. Ref.\cite{tzitrin2021fault} includes a full description of both $p$ and $q$ basis measurements ($q$ basis are required for logical gates) and for inclusion of $p$-squeezed states replacing some GKP states. Here we include description of the process for determining the error on $p$-basis measurements as these are the only basis required the threshold simulations we perform.

The macronode correction includes a squeezing gate and $X$ and $Z$ corrections. For a $p$-basis measurement on the central mode resulting in a homodyne value $m_\mathrm{central}$, the squeezing gate maps $p \rightarrow 2p$. This maps the GKP effective squeezing $\Delta \rightarrow 2\Delta$, resulting in an error probability of $p^\mathrm{central}_\mathrm{err}=\left(2m_\mathrm{central},2\Delta_\mathrm{central}\right)$. The $X$ corrections have no effect on the p-basis measurements so we do not consider them here. The $Z$ corrections are all the sum (or difference) or two independent normally distributed variables. This results in distribution with a variance that is the sum of the two normal distributions. This results in a error probability $p^\mathrm{corr}_i=\left(m_j \pm m_k,\sqrt{\Delta^2_j + \Delta^2_k}\right)$, where modes $j$ and $k$ are given by the specific correction, as outlined in Sec.~\ref{sec:macronode}. In the case where a correction is $0$, that is the $i^\mathrm{th}$ macronode is connected to the central mode of the neighboring macronode,  $p^\mathrm{corr}_\mathrm{err} = 0$, as shown in Fig.~\ref{fig:dict_protocol}(c).

\subsection{Logical error simulation method}\label{app:sim_method}
We follow a similar method to Ref.~\cite{Bourassa2021blueprintscalable} where we track the noise on each physical GKP mode. We start by assigning a separate noise variances in the $p$ and $q$ quadratures for every mode. This variance is either an abstract quantity defining the general noise channel in Eq.~\ref{eq:gauss_disp_channel} or the result from circuit-level PhANTM, squeezing, and breeding simulations. To determine the noise on the homodyne values, we propagate these variances using the symplectic representation of the beamsplitter network required to create the RHG lattice. The symplectic form, $\hat{S}$, transforms the quadrature values as $\textbf{x}^\prime = \hat{S}\textbf{x}$, where 
$\textbf{x} = \left(q_1,q_2,...q_N,p_1,p_2,... p_N\right)$. The noise variances are then transformed as 
\begin{equation}
    \Sigma_\mathrm{out} = \hat{S}^T \Sigma_\mathrm{in} \hat{S},
\end{equation}
\begin{figure*}[t!]
    \centering
    \includegraphics[width=1\linewidth]{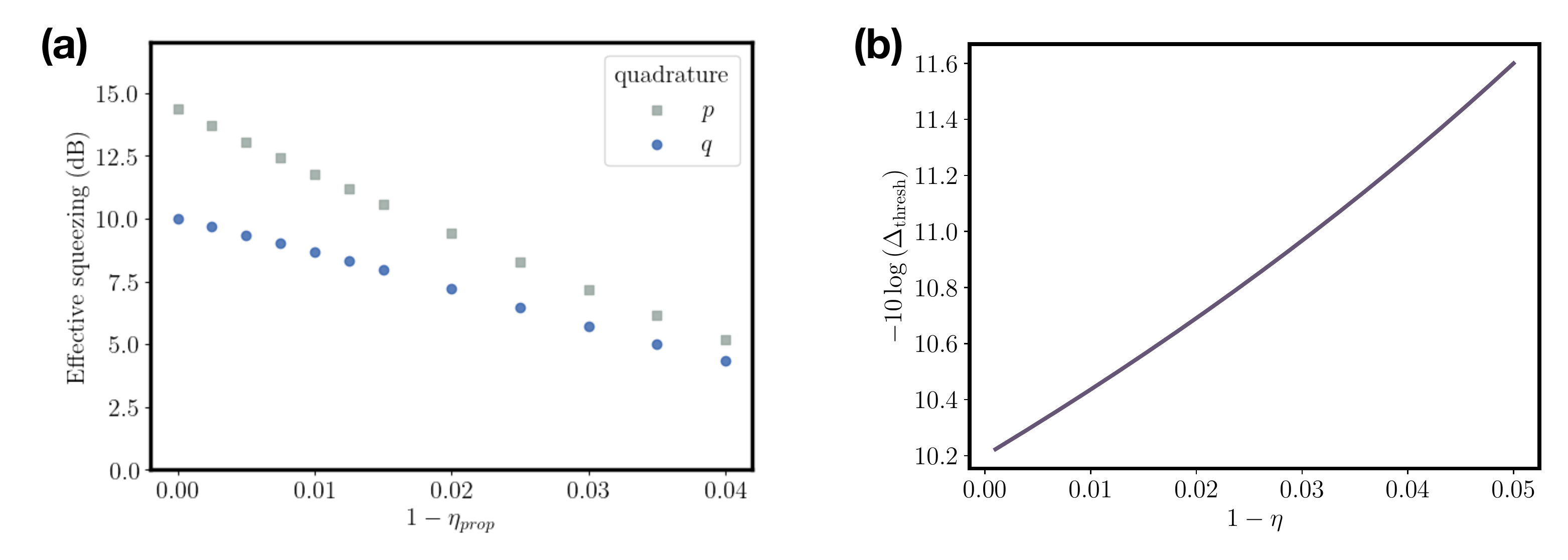}
    \caption{a) Effective squeezing in $p$ (square) and $q$ (circle) quadratures as a function of $1-\eta_{hd}$ for different cat state squeezing values. Only one type of squeezing is shown for the $p$-quadrature,  as for the other squeezing values the curves overlap with the one already on the plot. b) Loss vs squeezing threshold. We show the GKP effective squeezing that results in a total squeezing of $10.2$ dB for given values of $\eta$.}
    \label{fig:loss_pipeline}
\end{figure*}
where $\Sigma$ is a $2N \times 2N$ diagonal noise matrix, where the first $N$ entries are $\Delta^2_{q_i}$ and the second $N$ entries are $\Delta^2_{p_i}$.  We can simulate the noisy GKP measurements by drawing samples from the multi-variate distribution centred at 0 with a covariance matrix given by $\Sigma_\mathrm{out}$. Note that to simplify the simulations, we sample from the diagonal elements of $\Sigma_\mathrm{out}$.

We can convert these noisy measurements of the physical modes into error probabilities for the RHG lattice using the process described in Appendix~\ref{app:inner_decoder}. 

For a code distance $d$, we construct a $d\times d \times d$ lattice with periodic boundary conditions in all dimensions. Error correction proceeds using the code's parity check matrix, a binary matrix with elements $H_{ij} = 1$ if qubit $i$ is involved in parity check $j$ and 0 otherwise. We can then use the probabilities calculated above to generate a binary noise array, with element $\epsilon_i = 1$ with probability $p_i$. An element equal to 1 corresponds to a flipped measurement outcome, e.g. a Pauli Z error on qubit $i$. The corresponding syndrome can be found by matrix multiplication, modulo 2, $s = H\epsilon$. To decode the syndrome we use a minimum weight perfect matching decoder implemented in python with PyMatching \cite{higgott2023sparse}. This aims to return a possible error chain consistent with both the syndrome and parity check matrix. As in previous works, \cite{noh2020fault,larsen2021fault,Bourassa2021blueprintscalable,fukui2021all}, we can utilize the continuous nature of our measurements to increase the success probability of the syndrome decoder. We convert the error probabilities into weights for the decoder as $w_i = -\log\left(p_i\right)$. A logical error corresponds to the case where the combination of the original error and the decoder output result in an error chain that spans the lattice. This can be determined by checking the number of intersections of the resultant error chain with the logical operator correlation surface of the lattice. An odd number of intersections corresponds to logical error. Repeating this process many times allows us to determine the logical error rate for a particular GKP effective squeezing and code distance. As the primal and dual lattices can be decoded independently, we only show results for the primal lattice. The procedure for the dual lattice is identical.

\subsection{Loss tolerance }\label{sec:loss}

As mentioned in the main text, the effects of losses on cat states generated from PhANTM have been explored in \cite{Eaton2022Phantm}. Here, we discuss losses in adaptive breeding and QEC stages of our architecture. 

As in PhANTM~\cite{Eaton2022Phantm}, we model losses applied on an input state $\rho_{in}$ in adaptive breeding by mixing the state with vacuum on a beam splitter with a transmission coefficient $\eta$ and tracing over the reflective mode of the beam splitter, arriving at the output density matrix $\rho_{out}$ from a loss channel given by
\begin{equation}
    \rho_{out} = \sum_{l=0}^\infty \mathcal{L}_l \rho_{in }\mathcal{L}_l^\dagger,
\end{equation}
where each $\mathcal{L}_l$ is a Kraus operator expressed as
\begin{equation}
    \mathcal{L}_l = \sqrt{\frac{(1-\eta)^l}{l!\eta!}}a^le^{\frac{1}{2}a^\dagger a \ln(\eta)},
\end{equation}
and $l$ is an index for the Fock basis. 
Since adaptive breeding is performed on cat states and momentum squeezed states embedded in a cluster, first order losses come from homodyne detections which are modeled by putting a lossy channel before detection.
Fig. \ref{fig:loss_pipeline}a shows the effective squeezing of the GKP states in the $p$ and $q$ quadratures as a function of homodyne loss after three breeding rounds. To isolate the effect of losses from other random variations in the adaptive breeding protocol, we use a set of eight identical cat states as input and vary the squeezing $r_c$ of the cat states (the amplitude of the cats is fixed by the number of breeding rounds). We observe that the effective squeezing in the $q$ quadrature is not sensitive to the losses up to $5$\% loss. The effective squeezing in the $p$ quadrature is more sensitive than that in the other quadrature, but changes  by less than $0.5$ dB up to $\eta_{hd}\approx 0.03$. 

For QEC, loss on a GKP state can be mapped to the Gaussian displacement channel by rescaling any homodyne values~\cite{fukui2021all}. For transmission $\eta$ (representing the propagation loss), this results in a GKP state with a with variance $\Delta^2_\text{loss} = \frac{1-\eta}{2\eta}$. If we have a GKP state with some initial squeezing defined by the variance $\Delta^2_\text{GKP}$, then the total variance, including the loss, is $\Delta^2_\text{total} = \Delta^2_\text{GKP} + \Delta^2_\text{loss}$.  
For the case where we have balanced squeezing in both quadratures, our Monte-Carlo simulations produced a squeezing threshold of $10.2$ dB. For uniform loss on each physical mode in the macronode lattice the variance on the noisy homodyne values is described by $\Delta^2_\text{total}$. We can use this to determine the squeezing threshold in the presence of finite loss by finding $\Delta^2_\text{GKP}$ that, along with a given transmission $\eta$, gives $-10\log\left(\Delta^2_\text{total}\right) = 10.2$ dB. Fig.~\ref{fig:loss_pipeline}b shows the result of this simulation, highlighting that the needed effective squeezing to be at the fault tolerance threshold in the presence of $3$\% loss is increased by less than $1$ dB compared to the ideal case without loss.
\end{document}